\documentclass[10pt,a4paper,twoside]{article}

\usepackage{a4}
\usepackage{amsfonts}
\usepackage{amssymb}
\usepackage{epsfig}
\usepackage{psfig}

\title{
\vspace{-2cm}
\begin{flushright} {\small{HD-THEP-00-17\\ HU-EP-00/20}
    \\ \vspace{-.3cm} hep-th/0003171}
\end{flushright}
\vspace{2cm}
{\bf Two-loop Feynman Diagrams in Yang-Mills 
Theory from Bosonic String Amplitudes}}
\author{} 
\date{}

\input{epsf.sty}
\def\axowidth{0.5 }
\def\axoscale{1.0 }
\def\axoxoff{0 }
\def\axoyoff{0 }

\def\Gluon(#1,#2)(#3,#4)#5#6{
\put(\axoxoff,\axoyoff){
}
}

\def\Photon(#1,#2)(#3,#4)#5#6{
\put(\axoxoff,\axoyoff){
}
}

\def\ZigZag(#1,#2)(#3,#4)#5#6{
\put(\axoxoff,\axoyoff){
}
}

\def\PhotonArc(#1,#2)(#3,#4,#5)#6#7{
\put(\axoxoff,\axoyoff){
}
}

\def\GlueArc(#1,#2)(#3,#4,#5)#6#7{
\put(\axoxoff,\axoyoff){
}
}

\def\ArrowArc(#1,#2)(#3,#4,#5){
\put(\axoxoff,\axoyoff){
}
}

\def\LongArrowArc(#1,#2)(#3,#4,#5){
\put(\axoxoff,\axoyoff){
}
}

\def\DashArrowArc(#1,#2)(#3,#4,#5)#6{
\put(\axoxoff,\axoyoff){
}
}

\def\ArrowArcn(#1,#2)(#3,#4,#5){
\put(\axoxoff,\axoyoff){
}
}

\def\LongArrowArcn(#1,#2)(#3,#4,#5){
\put(\axoxoff,\axoyoff){
}
}

\def\DashArrowArcn(#1,#2)(#3,#4,#5)#6{
\put(\axoxoff,\axoyoff){
}
}

\def\ArrowLine(#1,#2)(#3,#4){
\put(\axoxoff,\axoyoff){
}
}

\def\LongArrow(#1,#2)(#3,#4){
\put(\axoxoff,\axoyoff){
}
}

\def\DashArrowLine(#1,#2)(#3,#4)#5{
\put(\axoxoff,\axoyoff){
}
}

\def\Line(#1,#2)(#3,#4){
\put(\axoxoff,\axoyoff){
}
}

\def\DashLine(#1,#2)(#3,#4)#5{
\put(\axoxoff,\axoyoff){
}
}

\def\CArc(#1,#2)(#3,#4,#5){
\put(\axoxoff,\axoyoff){
}
}

\def\DashCArc(#1,#2)(#3,#4,#5)#6{
\put(\axoxoff,\axoyoff){
}
}

\def\Vertex(#1,#2)#3{
\put(\axoxoff,\axoyoff){
}
}

\def\Text(#1,#2)[#3]#4{
\dimen0=\axoxoff \unitlength
\dimen1=\axoyoff \unitlength
\advance\dimen0 by #1 \unitlength
\advance\dimen1 by #2 \unitlength
\makeatletter
\@\raise\dimen1\hbox to\@{\kern\dimen0 \makebox(0,0)[#3]{#4}\hss}
\ignorespaces
\makeatother
}

\def\BCirc(#1,#2)#3{
\put(\axoxoff,\axoyoff){
}
}

\def\GCirc(#1,#2)#3#4{
\put(\axoxoff,\axoyoff){
}
}

\def\EBox(#1,#2)(#3,#4){
\put(\axoxoff,\axoyoff){
}
}

\def\BBox(#1,#2)(#3,#4){
\put(\axoxoff,\axoyoff){
}
}

\def\GBox(#1,#2)(#3,#4)#5{
\put(\axoxoff,\axoyoff){
}
}

\def\Boxc(#1,#2)(#3,#4){
\put(\axoxoff,\axoyoff){
}
}

\def\BBoxc(#1,#2)(#3,#4){
\put(\axoxoff,\axoyoff){
}
}

\def\GBoxc(#1,#2)(#3,#4)#5{
\put(\axoxoff,\axoyoff){
}
}

\def\SetOffset(#1,#2){\def\axoxoff{#1 } \def\axoyoff{#2 }}

\def\fsize{10 }

\def\PText(#1,#2)(#3)[#4]#5{
\ifx#4 lt{\def\fmode{0 }}\else{
\ifx#4 tl{\def\fmode{0 }}\else{
\ifx#4 lb{\def\fmode{2 }}\else{
\ifx#4 bl{\def\fmode{2 }}\else{
\ifx#4 l{\def\fmode{1 }}\else{
\ifx#4 rt{\def\fmode{6 }}\else{
\ifx#4 tr{\def\fmode{6 }}\else{
\ifx#4 rb{\def\fmode{8 }}\else{
\ifx#4 br{\def\fmode{8 }}\else{
\ifx#4 r{\def\fmode{7 }}\else{
\ifx#4 t{\def\fmode{3 }}\else{
\ifx#4 b{\def\fmode{5 }}\else{ \def\fmode{4 } }\fi
}\fi}\fi}\fi}\fi}\fi}\fi}\fi}\fi}\fi}\fi}\fi
\put(#1,#2){\makebox(0,0)[]{\special{"/pfont findfont /fsize scale setfont
\axoxoff \axoyoff #3 \fmode \fsize (#5) ptext }}}  }

\def\GOval(#1,#2)(#3,#4)(#5)#6{
\put(\axoxoff,\axoyoff){
}
}

\def\Oval(#1,#2)(#3,#4)(#5){
\put(\axoxoff,\axoyoff){
}
}

\let\eind=]

\def\kromme(#1,#2)#3{#1 #2 \ifx #3\eind\else\expandafter\kromme\fi#3}

\def\LogAxis(#1,#2)(#3,#4)(#5,#6,#7,#8){
\put(\axoxoff,\axoyoff){
}
}

\def\LinAxis(#1,#2)(#3,#4)(#5,#6,#7,#8,#9){
\put(\axoxoff,\axoyoff){
}
}

\newcommand{\beq}{\begin{equation}}
\newcommand{\eeq}{\end{equation}}
\newcommand{\beqn}{\begin{eqnarray}}
\newcommand{\eeqn}{\end{eqnarray}}

\newcommand{\wf}{World Line Formalism}
\newcommand{\gf}{Green's function}
\newcommand{\poa}{P_2(t_1,t_2,t_3,t_4,t_5)}

\newcommand{\non}{\nonumber \\}

\newcommand{\tr}{\mbox{tr}}

\newcommand{\dt}{{\mbox{det}}}

\newcommand{\al}{\alpha^\prime}
\newcommand{\alq}{\alpha^{\prime 2}}

\newcommand{\eps}{\epsilon}

\begin{document} 

\maketitle

\begin{center}
{\bf Boris K\"ors}\footnote{Email: Koers@Physik.HU-Berlin.De}$^{,a}$
and 
{\bf Michael G. Schmidt}\footnote{Email: M.G.Schmidt@ThPhys.Uni-Heidelberg.De}$^{,b}$ \\

\vspace{0.5cm}
$^a$Humboldt Universit\"at zu Berlin \\
{\small{Institut f\"ur Physik, Invalidenstr. 110, 10115 Berlin, Germany}} \\
\vspace{0.5cm}
$^b$Ruprecht-Karls-Universit\"at Heidelberg \\
{\small{Institut f\"ur Theoretische Physik, 
Philosophenweg 16, 69120 Heidelberg, Germany}} \\
\end{center}

\vspace{3cm}

\begin{abstract}
We present intermediate results of an ongoing investigation which 
attempts a generalization 
of the well known one-loop Bern Kosower rules of Yang-Mills theory 
to higher loop orders. We set up a general procedure to extract the field 
theoretical limit of bosonic open string diagrams, based on the sewing 
construction of higher loop world sheets. It is tested with one- and two-loop 
scalar field theory, as well as one-loop and two-loop vacuum Yang-Mills diagrams, 
reproducing earlier results. It is then applied to two-loop two-point Yang-Mills 
diagrams in order 
to extract universal renormalization coefficients that can be compared to field theory. 
While developing numerous technical tools to compute the relevant contributions, we 
hit upon important conceptual questions: Do string diagrams reproduce Yang-Mills 
Feynman diagrams in a certain preferred gauge? Do they employ a certain preferred 
renormalization scheme? Are four gluon vertices related to three gluon vertices? 
Unfortunately, our investigations remained inconclusive 
up to now. 
\end{abstract}

\thispagestyle{empty}
\clearpage

\tableofcontents

\section{Introduction}

Achievements developed by Bern and Kosower \cite{BK,BK2,BK3} allow to deduce the 
contributions of a large number of one-loop Feynman diagrams in Yang-Mills 
gauge theory from a single string scattering diagram \cite{MTZ}. 
This was first noticed by 
analyzing amplitudes of some heterotic string model, but later on it was realized 
that only the bosonic degrees of freedom were relevant in the appropriate field 
theoretical limit. In this sense one can say that the bosonic string amplitude adds up 
implicitly all particle diagrams of a given loop order. To extract these contributions 
one has to get rid of the massive modes of the string as well as its tachyonic 
excitation. If one were able to compute the corresponding zero Regge slope 
limit of the entire amplitude also in higher loop orders, 
a tremendous simplification of the computation of loop corrections in gauge 
theories might follow, 
which could be of great impact in perturbative techniques. It was further 
noticed that the string theoretical input to this method can be reduced to the 
quantum mechanics of particles moving on some world line which is reminiscent 
of the string world sheet at infinite string tension. Thus the \wf\ 
\cite{STR,SS3,CEF} was established and its relation to string theory explored 
\cite{RS1,RS2}. During this investigation it was also realized that the tachyonic 
mode of the string could be employed to calculate scattering amplitudes of scalar 
field theory despite its unphysical mass. As well a conclusive first quantized 
treatment of QED 
could be defined \cite{SSX,SS1} 
without regarding the string theoretical origin of the formalism 
any more. On the other hand, the foremost challenge of the entire enterprise still 
remains unsolved, which is the generalization of the Bern Kosower rules of 
Yang-Mills theory to higher loop orders. In fact the World Line formalism did allow 
higher order computations in QED and scalar field theory 
\cite{CEF,FRSS,KS}. It was also 
possible to extract such information directly from string 
diagrams \cite{DV6,DV7,MP,FMR}. Any attempt to master two-loop Yang-Mills theory 
starting from string diagrams remained inconclusive up to now. In \cite{SaS,SSZ} 
a worldine approach to two-loop Yang-Mills theory was proposed, whose connection to 
string theory has not yet been explored. \\

In this article which is an extension of relevant parts of \cite{HEU,KO} we try to 
define such a generalization in the spirit of the mentioned earlier works. Thus 
we first review parts of these and deduce 
a general procedure how to project the string amplitude onto the massless vector 
boson mode. Our method attempts to unify the formerly 
employed ones, reproducing all their results and making the two-loop extension 
straightforward at the first glance. We are able to present an exact computation 
of all contributions of the two-loop two gluon string amplitude, that are relevant to 
obtain the two-loop coefficient of the Yang-Mills $\beta$ function.  
But in the end we shall uncover 
two severe difficulties which we are unable to deal with, preventing us from 
identifying two-loop Bern-Kosower rules. One of these is the unanswered question 
of how the gauge choice of field theory enters into the string amplitude. We 
demonstrate that a comparison of Feynman diagrams of a particular ``topology'' to 
their string counterpart does not allow any of the so-called 
covariant background gauges. 
At the one-loop level the diagrams were reproduced in the particular Feynman 
background gauge and this was therefore supposed to be the preferred  
gauge in which the string amplitudes naturally 
appear in the field theory limit. The second open question is, how the string diagrams 
deal with renormalization. While string theory contains a scale at which the 
massive string modes cut off the ``divergencies'' of local field theories, these 
require renormalization. This, for instance, calls for the presence of counter term 
insertions contributing at the two-loop level of perturbation theory, 
a procedure which in general will depend on the renormalization scheme chosen. 
In which manner these contributions are included in string diagrams is rather  
obscure. Using background techniques, like we do in this paper, one also has to 
inspect IR regularization, before one can compare with the conventional $\beta$ function
\cite{SSZ}. \\ 

A further technical obstruction is our present inability to give a consistent treatment 
of four-gluon vertices. These do not generically occurr in string diagrams which 
are built up by sewing together three-point string vertices. In the limit when 
the geometry of the string world sheet tends to a diagram involving four-point 
vertices some of the moduli are frozen and their integrations have to be 
removed, which leaves one with divergencies that cannot be explained or 
regularized in an obvious manner. Thus our work unfortunately remains 
inconclusive in these respects but still allows some insight into the problems that 
prevent a further progress so far.\\    

The work on these topics is still in progress and we would like to express our 
appreciation for the discussion on this paper 
with P. Di Vecchia, A. Lerda, R. Marotta and R. Russo 
in fall 1998 in Copenhagen.

\section{The field theoretical limit of string amplitudes}
\label{fg}

The observation that field theoretical amplitudes can be recovered from string 
theory traces back to Scherk \cite{SCHE}, who found that expressions obtained 
from the dual operator formalism coincide precisely with field theortical result 
for tree-level and one-loop diagrams of $\Phi^3$ theory in the limit of vanishing 
Regge slope. One only has to introduce a relation between the string and 
field theoretical couplings, the whole kinematics of Feynman diagrams follows 
automatically.\\

From the string theoretical point of view one expects that in this limit the massless 
modes of the string particle spectrum form an effective low energy field theory. In 
the simple bosonic model these particle-like states are the scalar 
tachyonic excitation and the massless vector boson of the spectrum of the bosonic 
string.  When the string tension goes to infinity, i.e. $\al \rightarrow 0$, the 
former will be identified as a divergent $1/\al$ contribution of the amplitude 
while the latter is given by its constant term. \\

We shall identify these divergent, respectively constant contributions by introducing 
proper time variables, playing the same role as Schwinger proper times (SPT) in 
field theory. These are introduced when rewriting the momentum integrals over the 
internal momenta as 
\beqn \label{impspt}
\int{\frac{d^dp}{(2\pi)^d} \frac{p_\mu p_\nu ...}{p^2+m^2} =
  \int_0^\infty{dt \int{\frac{d^dp}{(2\pi)^d} p_\mu p_\nu...\ 
      e^{-t(p^2+m^2)} }} } .
\eeqn
One can now perform the Gaussian momentum integrations to obtain a SPT integral. 
This gives the type of integral which is naturally derived from string diagrams, by the 
conservation of difficulties it is technically not easier to solve than the 
Feynman momentum integral itself. The integration region is singled 
out by reducing the moduli space which has to be 
integrated over in the amplitude to that small region where the SPT variables stay 
finite \cite{BCFR}. The SPT variables are defined in string theory by identifying 
\beqn
\delta \tau (L_0-a) =  \delta \tau \al \left( p^2 +m^2 \right) 
\equiv \delta t \left( p^2 +m^2 \right) ,
\eeqn
or
\beqn
t =  \al \tau +t_0 =\al \ln \vert z \vert +t_0.
\eeqn
Logarithms of moduli correspond to proper time variables in the field theory. 
This reveals a truly geometrical interpretation of Feynman diagrams, which in 
this picture represent particles whose propagation is parametrized by proper time 
variables. The length of a particular propagator is defined by the fact that 
$x^{-L_0}$, $x$ ranging from $0$ to $1$, is the operator that propagates 
the external states in the string diagrams along the boundary of the open world 
sheet. It leads to
\beqn \label{sptmod}
t =-\al \ln (x)\quad \mbox{and}\quad x = e^{-t/\al}.
\eeqn
We thus have found means to extract the tachyonic and vector contribution from the 
string amplitude in terms of different powers in the moduli of the world sheet. The 
tachyon part of the amplitude is exponentially divergent in $\al$, or 
proportional to $1/x$, the gluon part is constant when $x \rightarrow 0$ and 
all massive states are exponentially suppressed. The full integration region 
in the moduli space is defined by proper times $t \in [0,\infty]$, which translates to 
$x \in [0,1]$. The insertion points of the external 
states are being integrated over all connected components of the world sheet 
boundary. On the other hand, from the point of view of field theory we could 
already be satisfied with $x\in [0,\eps]$. \\

We shall now set up a systematic three step procedure to extract the field 
theoretical contribution from string amplitudes following the sewing procedure 
in the form of \cite{DV6,MP}. We first 
replace moduli by proper time variables according to (\ref{sptmod}) and then, 
secondly, eliminate all terms proportional to higher powers of moduli, keeping 
only those which are constant for gluon diagrams or the divergent parts, 
proportional to inverse powers of moduli, for scalar field theory. This method 
does to the present stage not enable to extract any mixed diagram, which involves  
couplings of two different types of particles. Finally we determine the 
integration region of the moduli. This program will be demonstrated to work 
on scalar theory and one-loop Yang-Mills diagrams, before we come to our main 
topic, the computation of two-loop Yang-Mills diagrams and the extraction of 
renormalization constants. \\

\section{Scalar theory}

The limit leading to the tachyonic mode of the string spectrum can be used to 
reproduce results for single Feynman diagrams as well as formulas for complete 
$n$-point functions, which are known from the World Line approach to field theory 
\cite{SS3,CEF,SS1}. All field theoretical results quoted will refer to the $\Phi^3$ 
theory defined by the Lagrangian
\beqn \label{skwirkung}
{\cal L} = \frac{1}{2} \partial_\mu \Phi \partial_\mu \Phi
-\frac{m^2}{2} \Phi^2 +\frac{\lambda}{3!} \Phi^3 .
\eeqn
We first briefly discuss the \wf, show how the \gf\ of the bosonic string reduces 
to the World Line \gf\ \cite{RS1}, and afterwards proceed to Feynman diagrams 
\cite{DV6,MP,FMR}.

\subsection{The bosonic \gf}
\label{fggf}

The \gf\ of the bosonic string can be written 
\beqn \label{strgf}
{\cal G}^{(h)}(z_i,z_j) = \ln \left(
\frac{E^{(h)}(z_i,z_j)}{\sqrt{V_i^\prime(0) V_j^\prime(0)}} \right)
- \frac{1}{2} \int_{z_i}^{z_j}{\omega^\mu} (2\pi\ \Im \left( \tau_{\mu
  \nu} \right) )^{-1} \int_{z_i}^{z_j}{\omega^\nu}   .
\eeqn
It is discussed in greater detail in the appendix \ref{schottky} which we refer to 
for the geometrical description of world sheets. While the full string 
amplitude does not depend on any change of coordinate, the tiny part which is 
extracted from it in the chosen limit does, and one has to take a choice which 
set of local coordinates $V_i(z)$ one wants to use. The tree-level \gf\ reduces to the 
inverse of the Laplacian on the flat plane:
\beqn \label{strgf0}
{\cal G}^{(0)}(z_1,z_2) = \ln \left| \frac{ z_1-z_2 }{\sqrt{ V_1^\prime(0)
    V_2^\prime(0) } } \right|.
\eeqn
The local coordinates in general are specified in a way that the insertion points 
of the external states are being integrated over only one component of the boundary, 
which corresponds in the field theoretical picture to selecting a particular 
loop for any external particle. If one then neglects the radii of the 
isometrical circles, the boundary component of a single loop becomes the interval 
between two fixed points on the real axis, both, circles and fixed points, 
referring to the Schottky map of this loop. We then use the explicit form
\beqn \label{lokko}
V_i^\prime(0) = \left| \frac{(z_i-\eta_j)(z_i-\xi_j)}{(\xi_j-\eta_j)}
\right|  
\eeqn 
for the local coordinates. They specify the loop $j$ that is generated by the 
Schottky maps whose fixed points are $\xi_j$ and $\eta_j$ and to which the 
external state $i$ is attached. In fact, they are just the inverse of the first 
abelian differentials on the world aheet
\beqn
V_i^\prime (0) =\left( \frac{\omega^j (z_i)}{dz_i} \right)^{-1}.
\eeqn  
On the tree-level world sheet one then defines SPT variables by 
$\ln (z_{1,2}) = -t_{1,2}/\al$ and find 
\beqn
{\cal G}^{(0)}(z_1,z_2) \sim \ln \left| \sqrt{\frac{z_1}{z_2}} \right| = \pm
\frac{(t_1-t_2)}{2\al},
\eeqn
depending on which $z_i$ is the larger one. This matches precisely to the tree-level 
World Line \gf.\\

Proceeding to loop-level needs an interpretation of the further moduli of the world 
sheet, which are the fixed points and multipliers of the generators of the Schottky 
group. The latter are connected to the length of that loop by 
\beqn
T =-\al \ln \vert k \vert ,
\eeqn
while the fixed points shall be treated later. We now have to expand the prime form 
and the period matrix of the world sheet into a power series in the multipliers of 
the Schottky group and neglect all terms which are suppressed when $\al \rightarrow 0$. 
The expansion is discussed thoroughly in appendix \ref{schottkyentw}. Using the 
results one finds: 
\beqn \label{gwgf} 
E(z_1,z_2) &=& (z_1-z_2) +o(k_i),\\
\left( 2\pi \Im \left( \tau_{\mu \nu} \right) \right) &=& -\delta_{\mu
    \nu}\ln (k_\mu) -(1-\delta_{\mu \nu}) \ln \left|
  \frac{(\eta_\mu-\eta_\nu)(\xi_\mu-\xi_\nu)}{(\eta_\mu-\xi_\nu)(\xi_\mu-\eta_\nu)} 
\right| +o(k_i), \non
\int_{z_1}^{z_2}{\omega^\mu} &=& \ln \left|
  \frac{(z_1-\eta_\mu)(z_2-\xi_\mu)}{(z_1-\xi_\mu)(z_2-\eta_\mu)} \right|
  +o(k_i). \nonumber
\eeqn
This simplifies for the one-loop result by choosing $\eta=0$ and $\xi=\infty$, so 
that we obtain:
\beqn
E^{(1)}(z_1,z_2) &\sim & (z_1-z_2),\\
\left( 2\pi \Im \left( \tau^{(1)}_{11} \right) \right)^{-1} &\sim &
  \frac{1}{\al T} , \non 
\left( \int_{z_1}^{z_2}{\omega^1}\right)^{(1)} &\sim & \frac{\vert t_1-t_2 \vert}
{\al}. \nonumber
\eeqn
This has to be substituted into (\ref{strgf}) and we recover the World Line one-loop 
\gf. A more complete discussion of the whole procedure is given in \cite{RS1} and 
also the case of higher loop orders is verified. This establishes the relation
\beqn
{\cal G}^{(g)}(z_1,z_2) \rightarrow \frac{1}{2\al} G^{(g)}_B(t_1,t_2)
\eeqn
between the \gf\ of the bosonic string and that of the scalar World Line formalism 
of field theory to any order in perturbation theory. \\
 
\begin{figure}[h]
\begin{picture}(450,200)

\CArc(160,10)(160,360,180)
\CArc(160,10)(10,360,180)
\Vertex(160,10)1
\Vertex(400,10)1

\CArc(200,10)(20,360,180)
\CArc(260,10)(20,360,180)
\Vertex(200,10)1
\Vertex(260,10)1

\LongArrow(280,10)(380,10)
\Line(0,10)(150,10)
\Line(170,10)(180,10)
\Line(220,10)(240,10)

\DashCArc(230,10)(50,360,180)1
\DashCArc(230,10)(10,360,180)1
\DashCArc(75,10)(75,360,180)1
\DashCArc(245,10)(75,360,180)1

\Text(150,3)[l]{$\eta_2=0$}
\Text(165,3)[l]{$\eta_1=\eta$}
\Text(201,3)[l]{$\xi_1=1$}
\Text(315,3)[l]{$\xi_2=\infty$}

\end{picture} 
\caption[]{The two-loop world sheet of the open string} \label{wsfixpfix}    
\end{figure}
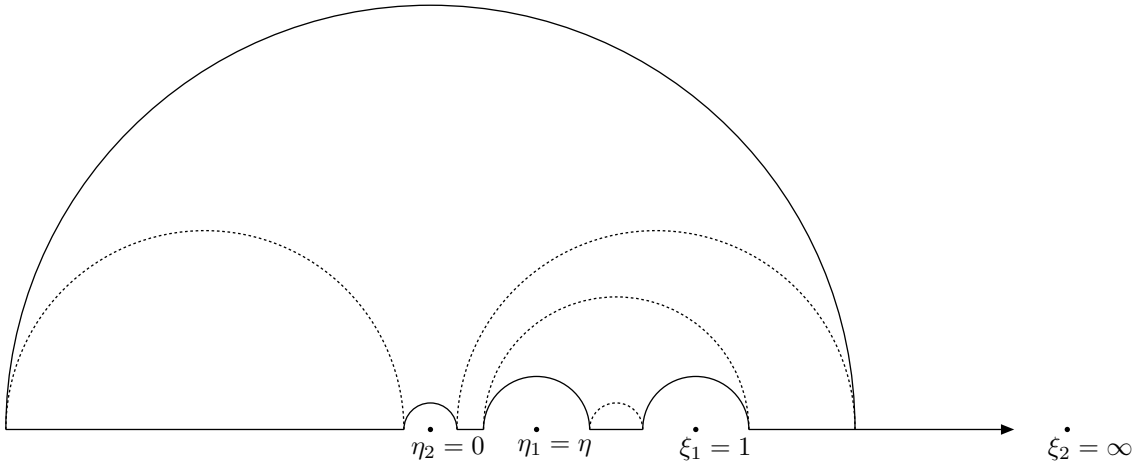

We now proceed to discuss the two-loop case. We always divide the 
integration measure of the integrals over moduli by the volume of the modular 
group of the Riemann surface by fixing three of the four fixed points to the values
\beqn \label{standkoord}
\eta_2=0, \quad \xi_2=\infty, \quad \xi_1=1
\eeqn
on the complex sphere, calling $\eta_1=\eta$. This standard choice is illustrated 
in figure \ref{wsfixpfix}. \\

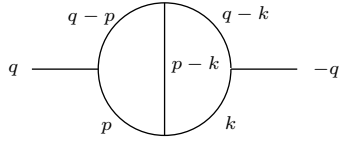
\begin{figure}[h]
\begin{picture}(400,100)

\CArc(200,30)(25,1,360)
\Line(150,30)(175,30)
\Line(225,30)(250,30)
\Line(200,5)(200,55)

\Text(200,32)[l]{\scriptsize{$p-k$}}
\Text(195,10)[l]{\scriptsize{$k$}}
\Text(123,8)[l]{\scriptsize{$p$}}
\Text(85,50)[l]{\scriptsize{$q-p$}}
\Text(118,51)[l]{\scriptsize{$q-k$}}
\Text(12,30)[l]{\scriptsize{$q$}}
\Text(102,30)[l]{\scriptsize{$-q$}}

\end{picture}
\caption[]{Scalar two-loop diagram}  
\label{skalgrzwlo}
\end{figure}

\begin{figure}[h]
\begin{picture}(400,70)

\LongArrow(50,30)(350,30)
\Line(50,30)(50,40)
\Line(100,30)(100,20)
\Line(150,30)(150,40)
\Line(200,30)(200,20)
\Line(250,30)(250,40)
\Vertex(370,30)1

\Text(40,48)[l]{$\eta_2=0$}
\Text(65,14)[l]{$z_2$}
\Text(90,48)[l]{$\eta_1=\eta$}
\Text(115,14)[l]{$z_1$}
\Text(140,48)[l]{$\xi_1=1$}
\Text(236,48)[l]{$\xi_2=\infty$}

\end{picture}
\caption[]{Parametrization of the world sheet}  
\label{wfparamabb}
\end{figure}
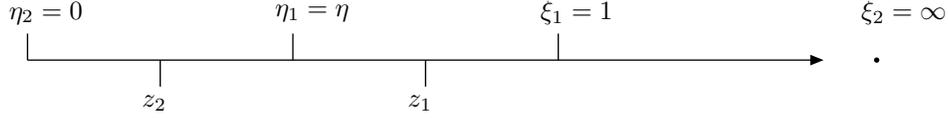

If we for instance intend to compute the diagram of figure \ref{skalgrzwlo} we can 
cut the world sheet as in figure \ref{wfparamabb}. 
The conformal invariance allows to cut in a way that $z_1,z_2,\eta < 1$. \\

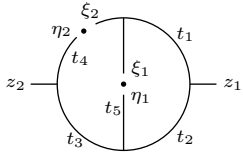
\begin{figure}[h]
\begin{picture}(400,100)

\CArc(200,30)(25,135,115)
\Line(165,30)(175,30)
\Line(225,30)(235,30)
\Line(200,5)(200,26)
\Line(200,34)(200,55)
\Vertex(200,30)1
\Vertex(185,50)1

\Text(200,25)[l]{\scriptsize{$\eta_1$}}
\Text(175,36)[l]{\scriptsize{$\xi_1$}}
\Text(167,10)[l]{\scriptsize{$t_2$}}
\Text(100,9)[l]{\scriptsize{$t_3$}}
\Text(69,50)[l]{\scriptsize{$\eta_2$}}
\Text(55,58)[l]{\scriptsize{$\xi_2$}}
\Text(66,49)[l]{\scriptsize{$t_1$}}
\Text(-24,30)[l]{\scriptsize{$z_2$}}
\Text(33,30)[l]{\scriptsize{$z_1$}}
\Text(-37,21)[l]{\scriptsize{$t_5$}}
\Text(-75,40)[l]{\scriptsize{$t_4$}}

\end{picture}
\caption[]{Sewing of the two-loop diagram}  
\label{abbsewzwlogr}
\end{figure}

We then read off 
\beqn
0=\eta_2 < z_2 < \eta_1=\eta < z_1 < \xi_1=1 < \xi_2=\infty 
\eeqn
and satisfy this relation by defining new moduli
$A_i \in [0,1]$ for $i=1,2,3$, which we substitute for the former ones: 
\beqn \label{standsew}
z_1 =A_1, \quad \eta = A_1A_2, \quad z_2= A_1A_2A_3 .
\eeqn
The $A_i$ are interpreted as sewing parameter which parametrize the length of 
the according pro\-pa\-ga\-tor. Their logarithms are translated into the proper 
time variables $t_i$ of the field theoretical diagrams. The loops support two more 
such SPT variables by their Schottky multipliers which are replaced by $t_4$ and 
$t_5$ so that the prescription altogether reads 
\beqn \label{sewa}
t_i &=& -\al \ln (A_i) , \quad \mbox{for}\ i=1,2,3,\\
t_5 &=& -\al \ln(k_1) -t_1-t_2, \non
t_4 &=& -\al \ln(k_2) -t_1-t_2-t_3 \nonumber .
\eeqn
This parametrization is depicted in figure \ref{abbsewzwlogr}. 
Any other logarithm occurring in the formula for the \gf\ will be translated 
analogously:
\beqn \label{logsub}
\al \ln(z_1-z_2) &=& -t_1+\al \ln (1-A_2A_3), \\
\al \ln(z_1z_2) &=& -2t_1-t_2-t_3, \non
\al \ln \left( \frac{z_1}{z_2} \right) &=& t_2+t_3, \non
\al \ln \left( \frac{(z_2-\eta)(z_1-1)}{(z_2-1)(z_1-\eta)} \right) &=&
-t_2 +\al \ln \left( \frac{(1-A_3)(1-A_1)}{(1-A_1A_2A_3)(1-A_2)} 
\right) \nonumber 
\eeqn 
and we finally find
\beqn
{\cal G}^{(2)}(z_1,z_2) &=& \ln (z_1-z_2) -\frac{1}{2} \ln (z_1 z_2)
-\frac{1}{\ln (k_1) \ln (k_2) - \ln^2(\eta)} \\
& & \times \left( \ln \left( \frac{(z_1-\eta)(z_2-1)}{(z_1-1)(z_2-\eta)} \right)
  \ln \left( \frac{z_1}{z_2} \right) \ln (\eta) - \ln \left(
  \frac{z_1}{z_2} \right) \ln (k_1) \right. \non 
& & \left. - \ln \left( \frac{(z_1-\eta)(z_2-1)}{(z_1-1)(z_2-\eta)}
\right) \ln (k_2) \right) +o(\alpha^{\prime 0}) \non 
&=& \frac{1}{2\al}
\frac{t_1t_2(t_3+t_4+t_5)+t_1t_3(t_4+t_5)+t_2t_4(t_3+t_5)+t_3t_4t_5}
{(t_1+t_2)(t_3+t_4+t_5)+(t_3+t_4)t_5}
+o(\alpha^{\prime 0}) . \nonumber  
\eeqn
This is precisely the result of the Gaussian integration which has to be performed 
in field theory for the particular combination of propagators which belongs to the 
diagram drawn in figure \ref{skalgrzwlo}:
\beqn
\int{\frac{d^dp d^dk}{(2\pi)^{2d}} \frac{1}{p^2 k^2 (q-k)^2 (p-q)^2 (k-p)^2} } = 
\int_0^\infty{\left( \prod_{i=1}^5{dt_i} \right)
  \left( (t_1+t_2)(t_3+t_4+t_5)+(t_3+t_4)t_5 \right) ^{-d/2}} \non 
\times\ \exp \left( -q^2
\frac{t_1t_2(t_3+t_4+t_5)+t_1t_3(t_4+t_5)+t_2t_4(t_3+t_5)+t_3t_4t_5}
{(t_1+t_2)(t_3+t_4+t_5)+(t_3+t_4)t_5}
\right) . \nonumber
\eeqn

\subsection{Scalar Feynman diagrams}

The relation between the string coupling $g_S$ and the field theoretical 
coupling $\lambda$ of the $\Phi^3$ theory from (\ref{skwirkung}) has to be chosen
\beqn
\lambda = 16 (2\al)^{(d-6)/4}g_S .
\eeqn
Furthermore a treatment of the integration measure and the normalization constants 
must be defined. The two-loop two-tachyon amplitude is constructed by inserting 
two scalar vertex operators into the path integral and it follows \cite{DV6}  
\beqn
{\cal A}_2^{(2)}(p,-p) = C_2 \left( 2g_S (2\al)^{(d-2)/4} \right) ^2
\int_\Gamma{\left[ dm \right]_2^2 \exp \left( 2\al p^2 {\cal
    G}^{(2)}(z_1,z_2) \right) } .
\eeqn
The nature of the normalization factor  
\beqn
C_2 \equiv \left( (2\pi)^2 \sqrt{2\al} \right) ^{-d} g_S^2
\eeqn
will also be discussed later on more thoroughly. The integration measure is given by: 
\beqn
\left[ dm\right] _2^2 &=& \frac{dk_1 dk_2 d\eta }{k_1^2 k_2^2
    (\eta-1)^2} \frac{dz_1 dz_2}{V_1^\prime(0) V_2^\prime(0)}
  (1-k_1)^2 (1-k_2)^2 \\
& & \times\ \dt^{-d/2} \left( -i\tau_{\mu \nu} \right)
  \prod_\beta{ \left( \prod_{n=1}^\infty{ \left( 
    1-k_\beta^n \right) ^{2-d} } \left( 1-k_\beta \right) ^{-2}
\right) }. \nonumber
\eeqn
The volume of the projective group has already been divided out by introducing the 
standard choice (\ref{standkoord}) of coordinates. For the local coordinates we use 
the usual $V_i^\prime(0) =z_i$. Next we substitute the sewing parameters from 
(\ref{standsew}) and find 
\beqn
\int_\Gamma{\frac{dk_1 dk_2 d\eta dz_1 dz_2}{k_1 k_2 \eta z_1 z_2}} &=& \int_0^1{
  \prod_{i=1}^3{ \left( \frac{dA_i}{A_i} \right) } \frac{dk_1 dk_2}{ k_1
    k_2} } = \left( \frac{1}{\al} \right) ^5 \int_0^\infty{\prod_{i=0}^5{dt_i}}, \\
(2\pi)^{-d} \dt^{-d/2} \left( -i\tau_{\mu \nu}  \right) &=& \left( \ln(k_1)
\ln(k_2) -\ln^2 (\eta) \right) ^{-d/2}  +o(k_1,k_2) \non
&=& \left( (t_1+t_2)(t_3+t_3+t_5)+(t_3+t_4)t_5 \right) ^{-d/2} +o(k_1,k_2). \nonumber
\eeqn
As we intend to project exclusively onto the tachyonic excitations we 
can employ their mass-shell condition 
\beqn
\al m^2 = -1
\eeqn
to define a particle mass for the scalar field. 
We then discard all terms proportional to powers of any moduli and retain a field 
theory diagram of a massive scalar field, while all other contributions are 
suppressed exponentially or, in the case of the gluon, stay finite when $\al
\rightarrow 0$. In regard of the prefactor we can omitt terms proportional to 
$k_1,k_2$ in the integrand 
\beqn
\prod_\beta{\left( \prod_{i=1}^\infty{ \left( 1-k_\beta^i
  \right) ^{2-d} } \left( 1-k_\beta \right) ^{-2} 
\right) } (1-k_1)^2 (1-k_2)^2 = 1+o(k_1,k_2)
\eeqn
and following (\ref{standsew}) and (\ref{sewa}) we get:
\beqn
\frac{\eta}{k_1 k_2} = \exp \left( -m^2 \sum_{i=1}^5{t_i} \right) .
\eeqn
The final result for the two-loop two-tachyon amplitude reads:
\beqn
{\cal A}_2^{(2)}(p^2) &=& \frac{\lambda^4}{2^9(4\pi)^d}
\int_0^\infty{\prod_{i=1}^5{\left( dt_i\ e^{-m^2t_i} \right) }} \left(
  (t_1+t_2)(t_3+t_3+t_5)+(t_3+t_4)t_5 \right) ^{-d/2} \\
& & \times\ \exp \left( -p^2
\frac{t_1t_2(t_3+t_4+t_5)+t_1t_3(t_4+t_5)+t_2t_4(t_3+t_5)+t_3t_4t_5}{(t_1+t_2)(t_3+t_4+t_5)+(t_3+t_4)t_5}
\right). \nonumber
\eeqn
It reproduces the field theoretical result for the Feynman diagrams from figure 
\ref{skalgrzwlo} including any combinatorial factors. In \cite{DV6} this method 
is also applied to reducible diagrams and the same positive result was confirmed. \\

This procedure not only works for $\Phi^3$ theory but can also be extended 
to scalar $\Phi^4$ theory \cite{MP,FMR} by a different matching of the coupling 
constants. A problem arises with the moduli or SPT variables of those 
propagators which shrink to zero length. Their integrations need to be regularized by 
hand, a method which is conceptually not very satisfactory. 
Still it can be performed to get contributions consistent with results for scalar field 
theories, while we will not be able to extrapolate the method to gauge theories. 
Another open question which one should be able to tackle already in 
two-loop scalar diagrams concerns the presence, or absence, of counterterm diagrams. 
It has not been investigated, how their contributions to the two-loop 
scattering are somehow implicitly contained in the string diagram. If not, the 
missing part could depend on the renormalization scheme.

\section{Pure Yang-Mills gauge theory}
\label{ym}

After having analyzed the scalar field theory in terms of \gf s and Feynman diagrams 
we shall now treat the most interesting case of Yang-Mills gauge theory. The 
challenging aim of our investigation is the generalization of the Bern-Kosower 
rules which allow a greatly simplified computation of one-loop $n$-point functions 
in pure gauge theories \cite{BK,BK2,BK3}. We shall present the means to calculate 
exactly all contributions of the two-point two-loop string amplitude that survive 
in the field theory limit. In this context we shall particularly 
have to point out the difficulties arising from the unknown relation between 
different choices for the local coordinate on the world sheet and different 
gauge choices in the field theory. Furthermore the problem of constructing four-gluon 
vertices in string theory will be left a riddle.\\

Starting point is the $n$-gluon $h$-loop amplitude of the bosonic string which is given 
by the expectation value of $n$ gluon vertex operators computed in 
the background of a bosonic string theory on a Genus $h$ Riemann surface. The 
contractions of the world sheet coordinates are again performed 
using the \gf\ from (\ref{strgf}). We briefly collect the constituents of this 
amplitude
\beqn \label{masterampl}
{\cal A}^{(h)}_{n}(p_1,...,p_n) &=& N^h \tr \left( \lambda^{a_1} \cdots \lambda^{a_n}
\right) C_h {\cal N}_0^n \int_{\Gamma}{\left[ dm \right] ^n_h} 
\prod_{i<j}^{n}{ \exp \left( 2\al p_i p_j {\cal G}^{(h)}(z_i,z_j)
\right) } \\ 
& & \times\ \exp \left( \sum_{i \not= j}^{n}{\sqrt{2\al} (p_j
  \epsilon_i) \partial_{z_i}{\cal G}^{(h)}(z_i,z_j)}+ \frac{1}{2}
\sum_{i \not= j}^{n}{(\epsilon_i \epsilon_j) \partial_{z_i}
  \partial_{z_j}{\cal G}^{(h)}(z_i,z_j)} \right)
_{\mbox{\scriptsize m.l.}} \nonumber .
\eeqn
The prefactor $N^h \tr \left( \lambda^{a_1} \cdots
\lambda^{a_n} \right)$ is the Chan Paton factor of the diagram with external gauge 
charges $\lambda^{a_i}$ of some $SU(N)$ gauge group, whose adjoint representation 
is normalized as follows:
\beqn
\tr \left( \lambda^{a_i} \lambda^{a_j} \right) = \frac{1}{2}
\delta_{a_i a_j} .
\eeqn
The $\eps_i$ are the polarization vectors of the external gluons and from the 
expansion of the exponential one only has to keep the terms that are multilinear 
in the $\eps_i$. 
The general integration measure can be written in terms of the Schottky parameters 
that span the moduli space $\Gamma$:
\beqn \label{mastermass}
\left[ dm \right] ^n_h &\equiv & \prod_{m=1}^{h}{\left( \frac{dk_m d\xi_m
    d\eta_m}{k_m^2(\xi_m-\eta_m)^2} (1-k_m)^2 \right) }
\prod_{m=1}^{n}{(dz_m)} \frac{1}{dV_{abc}} \\
& & \times\ \left( \det(-i\tau_{\mu \nu}) \right) ^{-d/2}
\prod_{\beta}{ \left( \prod_{m=1}^{\infty}{(1-k_\beta^m)^{-d}}
  \prod_{m=2}^{\infty}{(1-k_\beta^m)^2} \right) }. \nonumber   
\eeqn
It contains the normalization constant involving the product over primary classes 
and the volume of the modular group $dV_{abc}$ in the denominator. The bosonic 
\gf\ is again be given by  
\beqn \label{mastergf}
{\cal G}^{(h)}(z_i,z_j) = \ln \left(
\frac{E^{(h)}(z_i,z_j)}{\sqrt{V_i^\prime(0) V_j^\prime(0)}} \right)
- \frac{1}{2} \int_{z_i}^{z_j}{\omega^\mu} (2\pi\ \Im \left( \tau_{\mu
  \nu} \right) )^{-1} \int_{z_i}^{z_j}{\omega^\nu}   
\eeqn
Note the difference in the depence on the local coordinates compared to \cite{DV7}, 
which has to be obeyed to get correct results in higher than one-loop order \cite{RS1}. 
We are discussing off-shell amplitudes and therefore do not 
demand the mass-shell and transversality relations 
\beqn
p^2 =0 , \quad p\eps =0 
\eeqn
to hold, although for the sake of brevity we often restrict ourselves to the term 
proportional to $(p_j \eps_i)(\eps_j p_i)$. The derivation of the normalization 
constant is found in \cite{DV7}. It uses tree-level three- and four-point 
amplitudes and factorization arguments to fix the dependence of the prefactor 
on the scale $\al$ and the dimensionless string coupling constant 
\beqn
g_S \equiv \frac{g}{2} (2\al)^{1-d/4} .
\eeqn
It is then found:
\beqn \label{masterkonstante}
{\cal N}_0 &=& 2 g_S (2\al)^{(d-6)/4}, \\
C_h &=& \left( (2\pi)^h \sqrt{2\al} \right) ^{-d} g_S^{2h-2}. \nonumber
\eeqn
The first factor reproduces the normalization of the Fourier transformations which 
come with loop calculations and also assigns the correct physical dimension to the 
amplitude, while the second carries the necessary power in the string coupling 
that is dictated by the Euler characteristic $\chi({\cal M})=2-2h$ 
of the Riemann surface. \\

\subsection{One-loop diagrams}
\label{ymeinloop}

Following the rules we have established when adressing scalar field theory we shall 
now proceed to gauge fields \cite{DV7}. We expand the integrand of the 
amplitude in the moduli, keep only the finite and non-vanishing part 
of the expansion when $\al \rightarrow 0$ and translate everything into the 
language of SPT variables. The results allow to confirm the Ward identities 
of the Feynman background gauge and to read off the coefficient of the 
Yang-Mills $\beta$ function from the wave function renormalization constant 
of the gauge field, which at one-loop order receives contributions only from 
a single Feynman diagram.\\

We first specify the formulas for the measure (\ref{mastermass}), the \gf\
(\ref{mastergf}) and the constants (\ref{masterkonstante}):
\beqn \label{einlostrgf}
\left[ dm \right]_1^n &=& \prod_{m=1}^n{\left( dz_m \right) }
  \frac{1}{dV_{abc}} \frac{dkd\eta d\xi}{k^2 (\eta-\xi)^2 } \left(
  -\frac{\ln(k)}{2\pi} \right) ^{-d/2} \prod_{m=1}^\infty{\left( 1-k^m
  \right) ^{2-d} } \\
&=& \prod_{m=2}^n{\left( dz_m \right) } \frac{dk}{k^2}  \left(
-\frac{\ln(k)}{2\pi} \right) ^{-d/2}  \prod_{m=1}^\infty{\left( 1-k^m
\right) ^{2-d} }  , \non
{\cal G}^{(1)}(z_i,z_j) &=& \ln (z_i-z_j) - \frac{1}{2} \ln (z_i z_j)
+ \frac{1}{2 \ln (k)} \ln^2 \left( \frac{z_i}{z_j} \right) \non
& & + \ln \left(
\prod_{m=1}^\infty{ \frac{ \left( 1-k^m \frac{z_i}{z_j} \right) \left(
    1-k^m \frac{z_j}{z_i} \right) }{ \left( 1- k^m \right) ^2 } }
\right) , \non
N_0^n C_h &=& \left( 2\pi \sqrt{2\al} \right)^{-d} (2\al g^2)^{n/2} .\nonumber 
\eeqn
We have used the usual fixed point choice $\eta=0$ and $\xi =\infty$ as well as 
the Lovelace type local coordinates $V_i^\prime(0) = z_i$. Further $z_2=1$ has been 
fixed by a projective transformation. \\

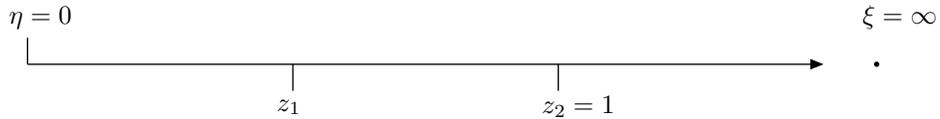
\begin{figure}[h]
\begin{picture}(400,70)

\LongArrow(50,30)(350,30)
\Line(50,30)(50,40)
\Line(150,30)(150,20)
\Line(250,30)(250,20)
\Vertex(370,30)1

\Text(40,48)[l]{$\eta=0$}
\Text(116,14)[l]{$z_1$}
\Text(191,14)[l]{$z_2=1$}
\Text(287,48)[l]{$\xi=\infty$}

\end{picture}
\caption[]{Parametrization of the one-loop two-point world sheet}  
\label{abbeinlozwpkt}
\end{figure}

\begin{figure}[h]
\begin{picture}(400,100)

\GlueArc(100,50)(25,1,180){2}{12}
\GlueArc(100,50)(25,180,360){2}{12}
\Gluon(50,50)(75,50){2}{4}
\Gluon(125,50)(150,50){2}{4}

\DashCArc(263,50)(25,1,360){3}
\Gluon(213,50)(238,50){2}{4}
\Gluon(288,50)(313,50){2}{4}

\Text(175,50)[l]{$+$}

\end{picture}
\caption[]{The one-loop Feynman diagrams that contribute to 
the Yang-Mills $\beta$ function}  
\label{abbgluoneinlo}
\end{figure}
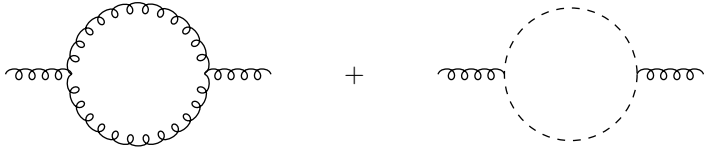

The only diagram can be parametrized as in figure \ref{abbeinlozwpkt} which leads 
to the integration region $\Gamma =\{ z_1\vert 0 < z_1 < 1\}$. Substituting the 
sewing parameter $z_1=A$ into  (\ref{einlostrgf}) and defining the proper times  
\beqn
T=-\al \ln(k), \quad t=-\al \ln (A), \quad [0,1] \rightarrow [\infty,0],
\eeqn
we find 
\beqn
\left[ dm \right]_1^2 &=& \frac{dk dz_1}{k^2}  \left(
-\frac{\ln(k)}{2\pi} \right) ^{-d/2}  \prod_{m=1}^\infty{\left( 1-k^m
\right) ^{2-d} }  \\
&=& \left( \frac{1}{\al} \right)^2 (2\pi)^{d/2} \frac{A T^{-d/2} dTdt}{k}
\left( 1+ (d-2)k +o \left( k^2 \right) \right), \non 
{\cal G}^{(1)}(z_1,z_2) &=& \ln (1-A) - \frac{1}{2} \ln (A) + \frac{1}{2 \ln (k)} 
\ln^2 \left( A \right) + \ln \left(
\prod_{m=1}^\infty{ \frac{ \left( 1-k^m /A \right) \left(
    1-A k^m \right) }{ \left( 1- k^m \right) ^2 } }
\right) \non
&=& \ln (1-A) +\frac{t}{2\al}  -\frac{t^2}{2\al T} +\sum_{n=1}^\infty{
  \left( \ln \left( 1-Ak^n \right) +\ln \left( 1-k^n/A \right) -2 \ln \left( 1-k^n
\right) \right) } .\nonumber    
\eeqn
In the same manner one can take the derivatives of the \gf. We now only have to keep 
the term in the integrand that is proportional to $A^0, k^0,
\alpha^{\prime 0}$, and thus finally find:
\beqn
{\cal A}^{(2)}_{2}(p^2) &=& N \tr \left( \lambda^{a_1} \lambda^{a_2} \right)
\frac{4g^2}{(4\pi)^{d/2}} \left( (\eps_2 p_1)(\eps_1 p_2) - \eps^2 p^2
\right) \int_0^\infty{ dT\ T^{-d/2} \int_0^\infty{ dt}} \\
& & \times  \left( \left(
    \frac{1}{2} - \frac{t}{T} \right)^2 (d-2) -2  +o(k,A) \right) \exp
    \left( -p^2 \left( t - \frac{t^2}{T} +o(\al,k,A) \right) \right)
    .\nonumber 
\eeqn
Rescaling the integration over $t$ by a factor $T$ leads to the result of \cite{DV7}. 
All integrations can easily be performed using some formulas given in appendix 
\ref{formeln} and the precise value that comes out of the computation of the 
sum of the two Feynman diagrams of Yang-Mills theory in the Feynman background gauge 
\cite{ABB}, drawn in \ref{abbgluoneinlo}, is recovered.

In particular from the wave function one-loop renormalization constant
\beqn
Z_A = 1+\frac{g^2N}{(4\pi)^2} \frac{11}{3} \frac{1}{\eps}
\eeqn
the correct coefficient of the Yang-Mills $\beta$ function \cite{ABB} is deduced. 
It is interesting to note that these notions obviously depend on the choice of local 
coordinates $V_i(z)$.\\

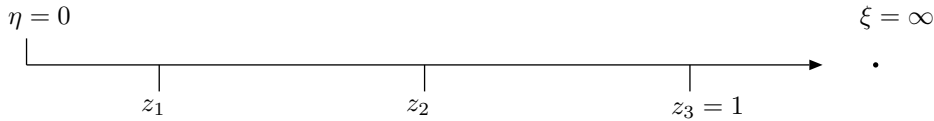
\begin{figure}[h]
\begin{picture}(400,70)

\LongArrow(50,30)(350,30)
\Line(50,30)(50,40)
\Line(100,30)(100,20)
\Line(200,30)(200,20)
\Line(300,30)(300,20)
\Vertex(370,30)1

\Text(40,48)[l]{$\eta=0$}
\Text(65,14)[l]{$z_1$}
\Text(140,14)[l]{$z_2$}
\Text(215,14)[l]{$z_3=1$}
\Text(261,48)[l]{$\xi=\infty$}

\end{picture}
\caption[]{Parametrization of the one-loop three-point world sheet}  
\label{abbeinlodreipkt}
\end{figure}

We now only briefly adress the one-loop three-gluon diagrams. This is the first and 
most simple example how to construct four-gluon vertices in string theory, which 
will be defined by extracting those regions in the moduli space where a propagator 
between two three-gluon vertices is short in terms of field theoretical proper time. 
We use the parametrization of figure \ref{abbeinlodreipkt}. \\

The integration region for the sewing parameters from (\ref{standsew}) will be as 
usually $[0,1]^3$. They are defined by 
\beqn
z_1 =A_1A_2, \quad z_2 =A_2
\eeqn
and lead to proper time variables
\beqn
t_i &=& -\al \ln (A_i), \quad \mbox{for}\ i=1,2, \\
t_3 &=& -\al \ln (k) -t_1-t_2 . \nonumber
\eeqn
We then get the sum of the two Feynman diagrams without four-gluon vertices by expanding 
the integrand and taking the appropriate limit. On the other hand, if we let $\al 
\rightarrow 0$ such that e.g. $A_1$ stays finite, we have $t_1 
\rightarrow 0$, although the insertion points $z_1$ and $z_2$ remain widely 
separated on the world sheet. The problem of this naive definition of four-gluon 
vertices in string amplitudes, the so-called pinching, is that it remains 
completely unclear what should happen to the free modulus $A_1$. In \cite{DV7} 
the term in the integrand, that stays finite when $A_1 \rightarrow 1$, is chosen 
to be relevant and the rest is being ignored. 
In fact, this method did allow to verify the relation between the 
three and four gluon diagrams deriving from the Ward identities 
of the Feyman background gauge for Yang-Mills theories and might thus be called 
heuristically successfull.\\

\subsection{Two-loop vacuum diagrams}
\label{zweiloopym}

\begin{figure}[h]
\begin{picture}(400,100)

\GlueArc(80,50)(25,90,270){2}{12}
\GlueArc(80,50)(25,-90,90){2}{12}
\Gluon(80,25)(80,75){2}{7}

\DashCArc(185,50)(25,1,360){3}
\Gluon(185,25)(185,75){2}{7}

\GlueArc(290,62)(10,90,450){2}{12}
\GlueArc(290,38)(10,-90,270){2}{12}

\Text(125,50)[l]{$+$}
\Text(97,10)[l]{(a)}
\Text(231,10)[l]{(b)}
\end{picture}
\caption[]{The three two-loop vaccum diagrams for Yang-Mills gauge theory}  
\label{abbgluonzweilovac}
\end{figure}
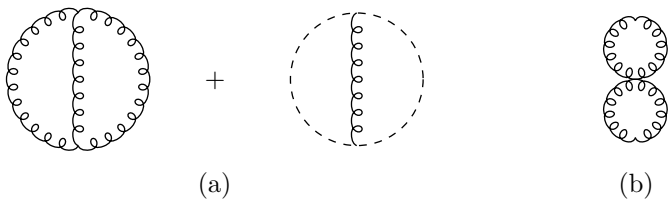

The most simple two-loop Feynman diagrams one can think of are the vacuum diagrams 
drawn in figure \ref{abbgluonzweilovac}. \\

They were discussed in \cite{MRU}. 
The generalization will not be straightforward as we shall 
have to deal with the announced difficulties concerning the diagram (b) with a 
four-gluon vertex. Another more fundamental question arises in the context of 
gauge invariance. The diagrams (a) and (b) clearly have to correspond to 
different regions in the moduli space of the string world sheet, simply because 
the number of field theoretical propagators is not the same, and therefore a 
different number of proper time variables has to be introduced. On the other 
hand their respective contributions in the field theoretical calculation will 
depend on the gauge one has chosen. One might expect that the statement which 
was true at one-loop level, that string contribution just come out in the Feynman 
background gauge, is still valid. The least one would like to require would be that added 
up they give a gauge invariant result that can be compared to field theory. 
Actually the question cannot be answered in the case of the vacuum 
diagrams, as their contributions vanish identically in dimensional regularization. 
This follows the principle that vacuum diagrams only give an irrelevant phase 
factor to the scattering matrix. So in the end there is no result to be extracted 
from this treatment, which we still undertake to point out the ealier mentioned 
problems that shall come up again whith the two-loop two-point function in the 
next chapter.\\

We now specialize all the relevant expressions in the amplitude, dropping 
the colour factor which is empty. To then work out the amlitude  
\beqn
{\cal A}_0^{(2)} = C_2 \int_\Gamma{\left[ dm \right]_2^0} =
  \frac{g^2}{4} \left( 2\pi \sqrt{2\al} \right) ^{-2d} (2\al )^2
  \int_\Gamma{\left[ dm \right]_2^0 }
\eeqn 
we only need the measure of the integration over the moduli of which some are fixed 
as in (\ref{standkoord}). It involves the period matrix
\beqn
2\pi \Im \left( \tau_{11} \right) &=& -\frac{\al
  \ln(k_1)}{\al}-\frac{2k_2(\eta-1)^2}{\eta} +o \left( k_1^2, k_2^2
\right), \\ 
2\pi \Im \left( \tau_{12} \right) &=& 2\pi \Im \left( \tau_{21}
\right) = -\frac{\al \ln(\eta)}{\al}
-\frac{2k_1 k_2(\eta+1)(\eta-1)^3}{\eta^2} +o \left( k_1^2, k_2^2
\right), \non 
2\pi \Im \left( \tau_{22} \right) &=& -\frac{\al
  \ln(k_2)}{\al}-\frac{2k_1(\eta-1)^2}{\eta} +o \left( k_1^2, k_2^2
\right), \nonumber 
\eeqn 
and its determinant 
\beqn \label{zwloperma}
\dt ^{-d/2}(-i\tau_{\mu \nu}) &=& (2\pi)^d \left( \frac{\al \ln(k_1)
  \al \ln(k_2)-\alq \ln ^2(\eta)}{\alq} \right) ^{-d/2} \\ 
& & \times \left( 1- \al d \left( \frac{(\eta-1)^2 (\al \ln(k_1) k_1
  +\al \ln(k_2) k_2)}{\eta (\al \ln(k_1) \al \ln(k_2)- \alq \ln
  ^2(\eta))} \right. \right. \non
& & \left. \left. -\frac{2 k_1 k_2 \al \ln(\eta)(\eta+1)(\eta-1)^3}{\eta^2
  (\al \ln(k_1) \al \ln(k_2) -\alq \ln ^2(\eta))} \right) +o(\alq,
k_1^2, k_2^2) \right) . \nonumber 
\eeqn
Together we find:
\beqn
\left[ dm \right]_2^0 &=& \frac{dk_1dk_2d\eta}{k_1^2 k_2^2 (1-\eta)^2} \dt
  ^{-d/2}(-i\tau_{\mu \nu}) \\
& & \times \prod_{\beta}{\left(
    \prod_{m=1}^{\infty}{(1-k_\beta^m)^{-d}}
    \prod_{m=2}^{\infty}{(1-k_\beta^m)^2} \right) } (1-k_1)^2(1-k_2)^2
  \non  
&=& \frac{dk_1 dk_2d\eta}{k_1^2 k_2^2} \dt ^{-d/2}(-i\tau_{\mu
    \nu}) \Bigg( 1+(d-2)(k_1+k_2) \non
& & + \left( (d-2)^2+\frac{d(1-\eta)^2(1+\eta^2)}{\eta^2} \right)
k_1k_2 +o\left( k_1^2, k_2^2 \right) \Bigg). \nonumber   
\eeqn
The derivation of these expressions is summarized in appendix \ref{schottky}. 
We have always added powers of $\al$ to get finite SPT variables in the end. 
After substituting for the integration over moduli 
\beqn
\int_\Gamma{\frac{dk_1 dk_2 d\eta}{k_1 k_2 \eta}} = \left(
\frac{1}{\al} \right)^3 \int_0^\infty{\prod_{i=1}^3{( dt_i )}} 
\eeqn
and regarding all relevant powers of the string tension, we are left with an over 
all factor $1/\al$. 
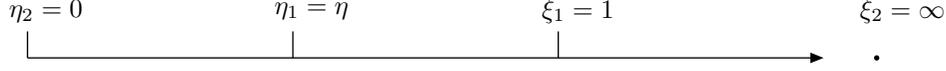
\begin{figure}[h]
\begin{picture}(400,70)

\LongArrow(50,30)(350,30)
\Line(50,30)(50,40)
\Line(150,30)(150,40)
\Line(250,30)(250,40)
\Vertex(370,30)1

\Text(40,48)[l]{$\eta_2=0$}
\Text(115,48)[l]{$\eta_1=\eta$}
\Text(191,48)[l]{$\xi_1=1$}
\Text(286,48)[l]{$\xi_2=\infty$}

\end{picture}
\caption[]{Parametrization of the two-loop vacuum world sheet}  
\label{abbzwlovacpara}
\end{figure}
The parametrization of figure \ref{abbzwlovacpara} using $\eta=A$ translates into 
\beqn
t_1=-\al \ln(A), \quad t_2 =-\al \ln(k_1) -t_1, \quad t_3=-\al
\ln(k_2) -t_1.
\eeqn
Now we have to extract the term proportional to $A^0, k^0, \al$ from the remaining 
integrand, which should belong to that part of the amplitude, i.e. that region in 
the moduli space, where all proper times stay finite. This corresponds to the sum of the \
two diagrams in (a) of figure \ref{abbgluonzweilovac}. We get 
\beqn
{\cal A}_0^{(2)}\vert_{\mbox{\scriptsize (a)}} = \frac{g^2}{(4\pi)^d}d(d-2)
\int_0^\infty{\prod_{i=1}^3{ (dt_i) } \frac{2t_1+t_2+t_3}{\left(
    t_1t_2+t_1t_3+t_2t_3 \right) ^{1+d/2} } },
\eeqn
which differs from the result in \cite{MRU} by the symmetrized integration region. 
The integral vanishes anyway in dimensional regularization.\\ 

Investigating the possible pinching contributions to diagram (b) of figure 
\ref{abbgluonzweilovac} we find the curious situation that three different limits 
of the string amplitude contribute to a single Feynman diagram. 
The regions where $t_2$ or
$t_3$ vanish, are defined by finite values for the moduli $k_1/A$ and $k_2/A$. 
We set $k_i\rightarrow Ak_i$, expand in $A$ and the other multiplier, keeping the 
finite part. The expansion in $\al$ has to be performed without any prefactor in 
$\al$ as the missing proper time integration leads to the missing of a factor $1/\al$. 
The integrand we get is finite when $k_i \rightarrow 1$ in both cases and the  
$k$-integration can be split off from the rest, giving the unique result 
\beqn
{\cal A}_0^{(2)}\vert_{\mbox{\scriptsize (b)}}^{t_{2,3}} =
\frac{g^2}{(4\pi)^d}  (d-2) \int{\frac{dk}{k^2}
  \int_0^\infty{\prod_{i=1}^2{ (dt_i) } \left( t_1 t_2 \right)^{-d/2} }} .
\eeqn 
The question what should happen to the integration over $k$ was answered in 
\cite{MRU} in the sense that one has to take an infinitesimal region around $k=1$ 
in the integrand and drop the integration. 
Even more problematic is the treatment of the integral 
over the free modulus in the case of the third possible pinching. This we get by 
having $t_1 \rightarrow 0$, i.e. $A$ finite. We do the same expansion as before:
\beqn 
{\cal A}_0^{(2)}\vert_{\mbox{\scriptsize (b)}}^{t_1} &=&
\frac{g^2}{(4\pi)^d}  \int{\frac{dA}{A^2(1-A)^2} \left( d(1+A^4)
    -2d(1+A^2)A +(d^2-2d+4)A^2 \right) } \\
& & \times \int_0^\infty{\prod_{i=1}^2{ (dt_i) } \left( t_1 t_2
\right)^{-d/2} }. \nonumber 
\eeqn 
The integral over $A$ is divergent at $0$ and $1$. In \cite{MRU} this was cured by 
a rather arbitrary zeta function regularization after expanding the integrand around 
$A=1$ and omitting the divergency at $A=0$ completely:
\beqn
\int_{1-\eps}^1{\frac{dA}{(1-A)^2}} \sim \zeta(0) = -\frac{1}{2} .
\eeqn
Similar divergent integrals have also been encountered in scalar $\Phi^4$ theory 
and similar methods, involving ``world sheet cut-offs'', have been used for 
regularization \cite{MP,FMR}. The result is
\beqn 
{\cal A}_0^{(2)}\vert_{\mbox{\scriptsize (b)}}^{t_1} =
- \frac{g^2}{2(4\pi)^d} (d-2)^2 \int_0^\infty{\prod_{i=1}^2{ (dt_i) }
  \left( t_1 t_2 \right)^{-d/2} }. 
\eeqn 
As we are unable to propose any reasonable alternatives, we do not intend to criticise 
these methods in detail. As mentioned a comparison of the contributions found from 
the string amplitude to field theory is impossible anyway, both are vanishing by 
definition. But we have to conclude, that certain divergent integrals over free 
moduli are appearing, if one tries to follow the strategies of the naive pinching 
procedures. These divergencies appear exactly in two different types which cover all the 
cases we investigated. If the proper time variable which is associated with the 
multiplier $k$ of a loop becomes small, we find an integrand 
finite when $k \rightarrow 1$ 
and the integration divergent of the kind $dk/k^2$, while any other sewing parameter 
$\eta$ leads to integrands that diverge like $d\eta/(\eta^2(1-\eta)^2)$. This shall be 
verified in the next chapter to be a generic feature of the pinching. The nature 
of the diverging integrals is very similar to those types of integrals that have 
to be performed when doing the trace over the Hilbert space of the intermediate 
string states when sewing together the world sheets of two strings, thus building 
up world sheets of higher Genus. In this sense we believe them to be related to the 
tachyon exchange in the shrinking propagator.

\subsection{Problems concerning the overlapping of isometric circles}

In this subsection we will explain a problem which arises when performing the 
$k\rightarrow 0$ limit of the string worldsheet. Up to now, and as well in the 
following, we completely ignore the requirement, that the circles cut out off the 
complex plane around the fixed points in the Schottky construction of the 
Riemann surface are not overlapping. Strictly speaking, this induces further 
restrictions on the moduli space, the allowed values of Schottky parameters, 
which in principle can be relevant 
also in the low energy limit. For all contributions to scalar field theory 
this apprantly has not been the case, because the naive sewing has been successful. 
But as we are unable in general to 
resolve the same task for Yang-Mills diagrams, this cannot be ruled out. 
In fact, in \cite{MRU} the prescription derived in \cite{DV7} was used, which implied 
a further restriction on the sewing parameters, as compared to our results above. 
But, luckily or not, 
the difference can be tracked down to an overall factor of 2 from symmetrizing 
the integrands in \cite{MRU}, which does not 
appear to be very significant in deciding whether the modification is necessary or not. 
Still it is worthwhile to be investigated in more detail. \\

In \cite{DV7} the Schottky parametrization was employed with circles different from the 
isometric ones. The latter are located at infinity, which makes 
the same treatment more difficult. Starting from the parametrization of diagram 
\ref{abbzwlovacpara} one adds the isometric circle around $\eta_1=\eta$ and $\xi_1=1$ 
and a circle of radius $\sqrt{k_2}$ around $\eta_2=0$ and its image with infinite radius 
but finite position at $1/\sqrt{k_2}$  
around $\xi_2=\infty$. Now $\eta$ can no longer vary freely between 0 and 1, but the two 
circles are further required not to intersect. This translates into the 
inequalities
\beqn
\sqrt{k_2} \leq \frac{\eta-\sqrt{k_1}}{1-\sqrt{k_1}} , 
\quad 0 \leq \frac{1-\sqrt{k_1}}{1+\sqrt{k_1}} \left( 1-\eta \right) .
\eeqn
The second inequality reduces to $\eta\leq 1$ for small $k_1$, whereas the first one is 
still a complicated relation between fixed points and multipliers, and several regions of 
their values will contribute. In \cite{DV7} the single inequality 
\beqn
0\leq \sqrt{k_2} \leq \sqrt{k_1} \leq \eta \leq 1
\eeqn
was used. It is not sufficient to solve the above requirements, but was 
enough to produce results consistent with scalar field theory expectations. The same 
inequality, translated to a different region of integration, was also used in \cite{MRU} 
for the vacuum diagrams of the Yang-Mills theory, 
but again we cannot judge their results to be decisive. 
Anyway, in more complicated diagrams, a systematic evaluation of all regions 
in the moduli space which respect the generalized version of the above inequalities 
may be necessary. One can easily convince oneself that for diagrams with external legs, 
the inequalities get much more complicated and a variety of distinctions arises. 
For some scalar $\Phi^4$ diagrams this task has been performed in \cite{FMR}. 
Also the methods we shall be employing to solve the unrestricted integrals do not allow 
a straightforward extension to this case, which leaves us with very little hope 
to be able to evaluate the analogous integrals. \\

\subsection{Two-loop diagrams involving external gluons}
\label{zwym}

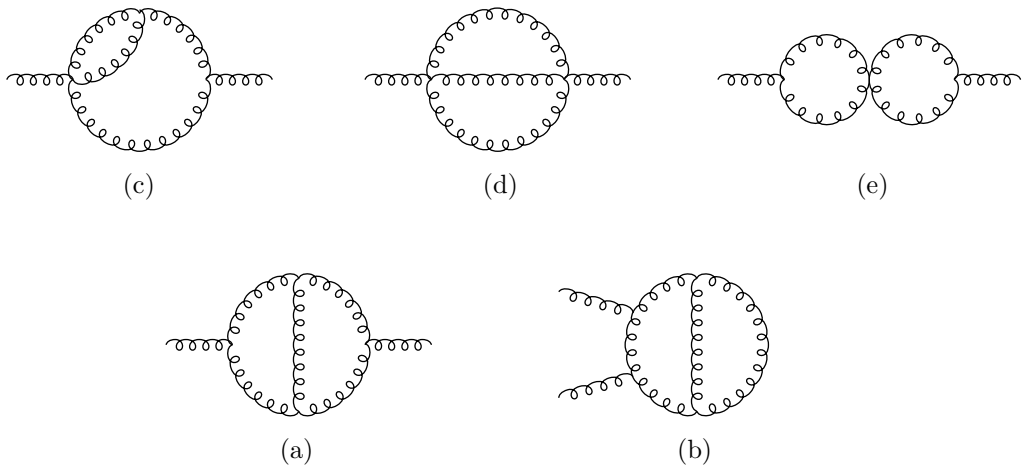
\begin{figure}[h]
\begin{picture}(400,200)

\GlueArc(110,50)(25,90,180){2}{6}
\GlueArc(110,50)(25,1,90){2}{6}
\GlueArc(110,50)(25,-180,-90){2}{6}
\GlueArc(110,50)(25,-90,1){2}{6}
\Gluon(60,50)(85,50){2}{4}
\Gluon(135,50)(160,50){2}{4}
\Gluon(110,25)(110,75){2}{8}

\GlueArc(260,50)(25,90,270){2}{12}
\GlueArc(260,50)(25,-90,90){2}{12}
\Gluon(208,70)(236,62){2}{4}
\Gluon(208,30)(236,38){2}{4}
\Gluon(260,25)(260,75){2}{8}

\GlueArc(50,150)(25,90,180){2}{6}
\GlueArc(50,150)(25,180,360){2}{12}
\GlueArc(50,150)(25,1,90){2}{6}
\Gluon(0,150)(25,150){2}{4}
\Gluon(75,150)(100,150){2}{4}
\GlueArc(25,175)(25,270,363){2}{6}

\GlueArc(185,150)(25,1,180){2}{12}
\GlueArc(185,150)(25,180,360){2}{12}
\Gluon(135,150)(160,150){2}{4}
\Gluon(210,150)(235,150){2}{4}
\Gluon(160,150)(210,150){2}{8}

\GlueArc(308,150)(15,-180,180){2}{12}
\GlueArc(342,150)(15,1,360){2}{12}
\Gluon(268,150)(293,150){2}{4}
\Gluon(357,150)(382,150){2}{4}

\Text(100,10)[l]{(a)}
\Text(225,10)[l]{(b)}
\Text(-10,110)[l]{(c)}
\Text(101,110)[l]{(d)}
\Text(217,110)[l]{(e)}

\end{picture}  
\caption[]{The gluonic two-loop contributions to the Yang-Mills two-point function}  
\label{abbgluonzweilo}
\end{figure}

We now come to our main topic, two-loop string diagrams with two external gluon vertex 
operators inserted. We shall again proceed along the lines we have established in the 
chapter on scalar theory and also we shall have to deal with the problems we 
already encountered in the previous two sections about Yang-Mills diagrams. 
This method should in principle allow to compute the two-loop cofficient 
of the Yang-Mills $\beta$ function 
\beqn \label{betaym}
\beta = -g \mu \frac{\partial Z_g}{\partial \mu} = -g \left( \beta_0 \left( \frac{g}{4\pi} \right)^2 + \beta_1
\left( \frac{g}{4\pi} \right)^4 +o\left( g^6 \right) \right) .
\eeqn
It can be extracted from the two point function, if the field theoretical gauge 
is any background type gauge. We shall admit an arbitrary covariant gauge of the 
gauge field fluctuations, 
which is still allowed after the background gauge is chosen.  In this 
kind of gauge the dependence of the charge renormalization constant on the subtraction 
scale can be deduced from the gauge field wave function 
renormalization constant alone. This in turn is given by the two point function:
\beqn
Z_A= 1+ \frac{\beta_0}{\eps} \left( \frac{g}{4\pi} \right)^2 +
\frac{\beta_1}{2\eps} \left( \frac{g}{4\pi} \right)^4 +o\left( g^6
\right) .
\eeqn 
The coefficients take the values \cite{CAS,JON}
\beqn 
\beta_0 = \frac{11}{3}, \quad \beta_1 = \frac{34}{3}.
\eeqn 
They are gauge invariant and therefore pose a good criterion to test the validity 
of the methods employed to higher loop order. All purely gluonic diagrams that 
contribute in field theory are drawn in figure \ref{abbgluonzweilo}. \\

The further diagrams involving ghost fields are obtained by substituting gluon loops 
by ghost loops. All such are listed in reference \cite{ABB}, where also their 
contributions in the Feynman background gauge are displayed. By the help of 
M. Peter from the Heidelberg University we are also able to discuss the contributions 
of these diagrams in an arbitrary covariant background gauge. In table 
\ref{tabpeter} they are summarized for $d=4-2\eps$, using the same letters for the 
diagrams as in reference \cite{ABB}, which are not to be confused with our 
notation of figure \ref{abbgluonzweilo}. We see that to the 
leading order in the $1/\eps$ expansion all diagrams are transverse, while this 
order necessarily adds up to zero. The next to leading order is not transverse 
diagram by diagram  but the sum of all contributions of course is. It remains 
unclear what we should expect to happen to the counter term diagrams, a 
problem that could not be investigated at one-loop order and has not been in 
scalar field theory.  
We should be able to notice their absence by a deviation 
of our result from the correct $\beta$ function coefficient by just their 
contribution. A very natural assumption seems, that we are dealing with a ``naked'' 
theory with string derived diagrams, where no counter terms are present. It is 
unclear, how the string diagrams, which orginally have a natural cut-off, 
implement renormalization. 
The explicit form of counter terms conventionally depends on the renormalization 
scheme employed, which would pose another puzzle. 
In other words: If the effects of the one-loop renormalization are included 
automatically, the result is universal, while if not, it will depend on the scheme. 
As we are unable to resolve the problems concerning 
four-gluon vertices, we shall not be in the position to unravel this question 
anyway. \\

\begin{table}[h]
\caption{Contributions of two-gluon 
two-loop diagrams in Yang-Mills theory in a general covariant 
background gauge} \label{tabpeter} 
\begin{center}
\begin{tabular}{lcc}
\vspace{0.2cm}
& Terms proportional to $\eps^2 p^2$ & Terms proportional to 
$(\eps p)^2$ \\
\vspace{0.2cm}
(a) & $ \left( \frac{1}{6} + \frac{\xi}{12} \right) \eps^{-2} + \left(
\frac{13}{12} + \frac{ 3\xi}{8} \right) \eps^{-1} $ & $ \left( 
\frac{1}{6} +\frac{\xi}{12} \right) \eps^{-2} + \left( 1+\frac{\xi}{3}
\right) \eps^{-1} $ \\ 
\vspace{0.2cm} 
(b) & $ \left( \frac{25}{6} + \frac{5\xi}{3} +\frac{\xi^2}{8} \right)
\eps^{-2} $ & $ \left(
\frac{25}{6} + \frac{5\xi}{3} + \frac{\xi^2}{8} \right) \eps^{-2} $ \\
\vspace{0.2cm}
& $  + \left( \frac{215}{12} + \frac{11\xi}{9} +
\frac{19\xi^2}{48} + \frac{\xi^3}{16} \right) \eps^{-1} $ & $  +
\left( \frac{55}{3} + \frac{67\xi}{72} +\frac{13\xi^2}{48}
  +\frac{\xi^3}{16} \right) \eps^{-1} $ \\ 
\vspace{0.2cm}
(c) & $ \left( \frac{1}{8} - \frac{\xi}{24} \right) \eps^{-1} $ &
$ -\frac{\xi}{24} \eps^{-1} $\\ 
\vspace{0.2cm}
(d) & $ \left( -\frac{9}{8} + \frac{7\xi}{8} - \frac{11\xi^2}{32}
+\frac{3\xi^3}{32} \right) \eps^{-1} $ & $ \left( \frac{\xi}{8}
-\frac{\xi^2}{4} + \frac{3\xi^3}{32} \right) \eps^{-1} $ \\  
\vspace{0.2cm}
(e) & $ \left( -6 + 3\xi - \frac{3\xi^2}{8} \right) \eps^{-2} 
 $ & $ \left( -6 + 3\xi - \frac{3\xi^2}{8} \right)
  \eps^{-2} $ \\  
\vspace{0.2cm}
& $ + \left( -24 +18\xi - \frac{9\xi^2}{2} + \frac{3\xi^3}{8} \right)
\eps^{-1} $ & $  + \left( -24 +18 \xi -\frac{9\xi^2}{2}  +
  \frac{3\xi^3}{8} \right) \eps^{-1} $ \\ 
\vspace{0.2cm}
(f) & $ \left( -\frac{1}{6} - \frac{\xi}{12} \right) \eps^{-2} +
\left( -\frac{13}{12} - \frac{3\xi}{8} \right) \eps^{-1} $ & $ \left(
  -\frac{1}{6} - \frac{\xi}{12} \right) \eps^{-2} + \left(
  -\frac{3}{4} - \frac{5\xi}{24} \right) \eps^{-1} $ \\
\vspace{0.2cm}
(g) & $ \left( -\frac{5}{24} + \frac{13\xi}{48} - \frac{\xi^2}{16}
\right) \eps^{-2} $ & $  \left( -\frac{5}{24} +
\frac{13\xi}{48} - \frac{\xi^2}{16} \right) \eps^{-2} $ \\
\vspace{0.2cm}
& $  + \left( -\frac{41}{48} +\frac{39\xi}{32} -
\frac{9\xi^2}{32} \right) \eps^{-1} $ & $  + \left(
-\frac{15}{16} +\frac{113\xi}{96} -\frac{9\xi^2}{32} \right) \eps^{-1}
$ \\
\vspace{0.2cm}
(h) & $ \left( -\frac{9}{8} - \frac{3\xi}{2} + \frac{9\xi^2}{32}
\right) \eps^{-2} $ & $
\left( -\frac{9}{8} - \frac{3\xi}{2} + \frac{9\xi^2}{32} \right)
  \eps^{-2} $ \\
\vspace{0.2cm}
& $  + \left( -\frac{57}{16} - \frac{19\xi}{4}
+\frac{125\xi^2}{64} - \frac{9\xi^3}{32} \right) \eps^{-1} $ & $
+ \left( -\frac{93}{16} - \frac{13\xi}{4} + \frac{113\xi^2}{64} -
  \frac{9\xi^3}{32} \right) \eps^{-1} $ \\
\vspace{0.2cm}
(i) & $ \left( \frac{1}{24} -\frac{\xi}{24} \right) \eps^{-2} + \left(
\frac{19}{48} -\frac{11\xi}{48} \right) \eps^{-1} $ & $  \left(
\frac{1}{24} -\frac{\xi}{24} \right) \eps^{-2} + \left( \frac{5}{16}
-\frac{5\xi}{24} \right) \eps^{-1} $ \\
\vspace{0.2cm}
(j) & $ \left( -\frac{1}{4} - \frac{\xi}{24} \right) \eps^{-2}  +
\left( -\frac{9}{8} - \frac{7\xi}{144} +\frac{\xi^2}{48} \right) \eps^{-1} $ & $
\left( -\frac{1}{4} - \frac{\xi}{24} \right) \eps^{-2}  + \left(
  -\frac{9}{8} - \frac{7\xi}{144} +\frac{\xi^2}{48} \right) \eps^{-1} $ \\
\vspace{0.2cm} 
(k) & $ \left( \frac{27}{8} -\frac{41\xi}{12} + \frac{\xi^2}{32}
\right) \eps^{-2} $ & $  \left(
\frac{27}{8} -\frac{41\xi}{12} + \frac{\xi^2}{32} \right) \eps^{-2} $ \\   
& $  + \left( \frac{233}{16} -\frac{121\xi}{9}
+\frac{203\xi^2}{64} -\frac{\xi^3}{2} \right) \eps^{-1} $ & $ + \left(
\frac{245}{16} -\frac{1999\xi}{144} +\frac{217\xi^2}{64}
-\frac{\xi^3}{2} \right) \eps^{-1} $ \\ 
\end{tabular}  
\end{center}
\end{table}

We now take (\ref{masterampl}) for $h=2$ and $n=2$ and use results from appendix 
\ref{schottky} employing the usual coordinate fixing (\ref{standkoord}) to get 
\beqn \label{zwlozwpktampl}
{\cal A}^{(2)}_{2}(p,-p) &=& N^2 \tr \left( \lambda^{a_1} \lambda^{a_2} \right)
C_2 {\cal N}_0^2 \int_{\Gamma}{\frac{dk_1 dk_2 d\eta dz_1 dz_2}{k_1^2
    k_2^2 (1-\eta)^2}(1-k_1)^2(1-k_2)^2} \\
& & \times\ \dt^{-d/2}(-i\tau_{\mu \nu}) \prod_{\beta}{\left(
  \prod_{m=1}^{\infty}{(1-k_\beta^m)^{-d}}
  \prod_{m=2}^{\infty}{(1-k_\beta^m)^2} \right) } \exp \left( -2\al
p^2 {\cal G}^{(2)}(z_1,z_2) \right)  \non  
& & \times \left( (2\al) (\epsilon_1 p_2)(p_2 \epsilon_1)
\partial_{z_{1}}{\cal G}^{(2)}(z_1,z_2) \partial_{z_{2}}{\cal
  G}^{(2)}(z_1,z_2)+ (\epsilon_1 \epsilon_2) \partial_{z_{1}}
\partial_{z_{2}}{\cal G}^{(2)}(z_1,z_2)  \right) ,\nonumber
\eeqn
where the expansion in the multipliers and $\al$ is partly done already. 
The determinant of 
the period matrix can be read off from (\ref{zwloperma}). The prime form is 
\beqn
E^{(2)}(z_1,z_2) &=& \frac{\al \ln (z_1-z_2)}{\al} - (z_1-z_2)^2
\left( \frac{k_1(\eta-1)^2}{(z_1-\eta)(z_2-\eta)(z_1-1)(z_2-1)} +
  \frac{k_2}{z_1z_2} \right. \\ 
& & + \left. \frac{k_1k_2(\eta-1)^2}{\eta^2} \left(
\frac{\eta^2+z_1z_2}{z_1z_2(z_1-1)(z_2-1)} +
\frac{\eta^2(1+z_1z_2)}{z_1z_2(z_1-\eta)(z_2-\eta)} \right) \right)
+o\left( k_1^2,k_2^2 \right) 
\nonumber 
\eeqn
and the abelian integrals follow
\beqn
\left( \int_{z_1}^{z_2}{\omega^1} \right) ^{(2)} &=& \frac{\al}{\al}
  \ln \left( \frac{(z_1-\eta)(z_2-1)}{(z_1-1)(z_2-\eta)} \right)
  -\frac{k_2(\eta-1)(z_2-z_1)(z_1z_2+\eta)}{\eta z_1z_2} \\ 
& & +\frac{k_1k_2(\eta-1)^3(z_1-z_2)}{\eta^2} \left(
\frac{\eta+1}{(z_1-1)(z_2-1)}
+\frac{\eta(\eta+1)}{(z_1-\eta)(z_2-\eta)} \right) +o \left( k_1^2,
k_2^2 \right), \non  
\left( \int_{z_1}^{z_2}{\omega^2} \right) ^{(2)} 
&=& \frac{\al}{\al} \ln \left( \frac{z_1}{z_2} \right)
+\frac{k_1(\eta-1)^2(z_1-z_2)}{\eta} \left( \frac{1}{(z_1-1)(z_2-1)}
+\frac{\eta}{(z_1-\eta)(z_2-\eta)} \right) \non 
& & +\frac{k_1k_2(\eta-1)^2(z_1-z_2)}{\eta^2} \left( \eta
+1+\frac{\eta(\eta+1)}{z_1z_2} \right) +o \left( k_1^2, k_2^2
\right) , \nonumber    
\eeqn
finally the inverse of the period matrix:
\beqn
\left( 2\pi \Im \left( \tau_{\mu \nu} \right) \right)^{-1} &=&  \dt
  ^{-1} \left( 2\pi \Im \left( \tau_{\mu \nu} \right) \right) \left(
  \begin{array}{ccc} \tau_{22} & -\tau_{12} \\ -\tau_{21} & \tau_{11}
  \end{array} \right), \\  
\dt ^{-1} \left( 2\pi \Im \left( \tau_{\mu \nu} \right) \right) &=&
\left( \frac{\al \ln(k_1) \al \ln(k_2)-\alq \ln ^2(\eta)}{\alq} \right) ^{-1} \non 
& & \times \Bigg( 1- \al \Bigg( \frac{2(\eta-1)^2(\al \ln(k_1) k_1+\al
  \ln(k_2) k_2)}{\eta (\al \ln(k_1) \al \ln(k_2)-\alq \ln ^2(\eta))}
\non 
& & -\frac{4k_1k_2 \al \ln(\eta) (\eta+1)(\eta-1)^3}{\eta^2 (\al \ln(k_1)
  \al \ln(k_2) -\alq \ln ^2(\eta))} \Bigg) +o\left( \alq, k_1^2,
k_2^2 \right) \Bigg). \nonumber  
\eeqn
For the product over the primary classes we use the expansion
\beqn 
& & \prod_{\beta}{\left(
  \prod_{n=1}^{\infty}{(1-k_\beta^n)^{-d}}
  \prod_{n=2}^{\infty}{(1-k_\beta^n)^2} \right)} (1-k_1)^2(1-k_2)^2 \\
&=& 1+(d-2)(k_1+k_2) + \bigg( (d-2)^2+ \frac{d (1-\eta)^2
  (1+\eta^2)}{\eta^2} \bigg) k_1k_2 + o(k_1^2,k_2^2) \nonumber
\eeqn
and the prefactor is 
\beqn
C_2 {\cal N}_0^2 = \frac{g^4}{4} \left( (2\pi) \sqrt{2\al} \right)
^{-2d} (2\al)^3.
\eeqn 
All expansions that have to be performed in the following are clearly impossible 
to be done by hand, so we regularly used the algebraic computer program MAPLE.
We now treat all the diagrams of figure \ref{abbgluonzweilo} one by one and keep 
track not to loose any contributions from the string amplitude. We start with (a), 
parametrize according to figure \ref{wfparamabb}, substitute sewing parameters as 
in (\ref{standsew}) and finally define proper time variables by (\ref{sewa}). 
The integration is transformed as 
\beqn
\int_\Gamma{\frac{dk_1dk_2d\eta dz_1dz_2}{k_1k_2\eta z_1z_2}} = \left(
\frac{1}{\al} \right) ^5 \int_0^\infty{\prod_{i=1}^5{(dt_i)}} .
\eeqn 
This corresponds to a sewing procedure that sewed the diagram together as depicted 
in figure \ref{abbsewzwlogr}. We next choose the insertion points of the external 
states to lie at the ``large'' loop, i.e. $\eta_2=0$ and $\xi_2=\infty$. The 
possibly astonishing fact, that the fixing of coordinates on the string world 
sheet leads to a distiction of different loops when looking for contributions 
to particular field theoretical Feynman diagrams has already been discussed in 
\cite{RS1,RS2}. It has been found how the local coordinates have to be chosen 
in order to get correct \gf s, which in our case requires us to take 
$V_i^\prime(0)=z_i$, as expected. The only way to have the external states on 
different loops is then given by the parametrization of figure \ref{wfparamabb}, 
except up to interchanging the external states, which does not give any different 
result, as has been checked explicitely in all cases. Using further (\ref{logsub}) 
and substitute everything into (\ref{zwlozwpktampl}) we find 
\beqn \label{zwloaint}
{\cal A}^{(2)}_{2} \left( p^2 \right) \vert _{\mbox{\scriptsize (a)}} &=& N^2
\delta_{a_1 a_2} \frac{g^4}{(4\pi)^d} \int_0^\infty{\prod_{i=1}^5{(dt_i)} } \\
& & \times \frac{\left( (\eps_1p_2)(\eps_2p_1) P_5^{(a)}(t_1,...,t_5,d)
     +\eps^2  P_{4^\prime}^{(a)} (t_1,...,t_5,d)\right) }{\left(
    (t_1+t_2)(t_3+t_4+t_5)+(t_3+t_4)t_5 \right)^{d/2+3} } \non
& & \times\ \exp  \left( -p^2
\frac{t_1t_2(t_3+t_4+t_5)+t_1t_3(t_4+t_5)+t_2t_4(t_3+t_5)+t_3t_4t_5}{(t_1+t_2)
(t_3+t_4+t_5)+(t_3+t_4)t_5}
\right) \non
&=&  N^2 \delta_{a_1 a_2}\left(  \frac{g}{4\pi} \right)^4 \left(
\frac{4\pi}{\sqrt{p^2}}\right)^{-\eps} \left( (\eps_1p_2)(\eps_2p_1) I_5^{(a)}(\eps) 
+ (\eps^2 p^2)
I_{4^\prime}^{(a)}(\eps) \right) . \nonumber 
\eeqn
The integrals $I^{(a)}$ must be computed in dimensional regularization. Each of the 
polynomials $P^{(a)}$ contains a couple of hundred terms and is therefore not written 
explicitely, instead we only display its degree as an index. In the second line we 
have extracted the momentum dependence in $d=4+\eps$ dimensions by rescaling 
the integration variables. \\

Power counting reveals that we have to deal at least with some logarithmic UV 
divergencies proportional to $1/\eps$ for small values of the SPT variables, 
whereas for $t_i \rightarrow \infty$ all integrals are IR finite. In a first 
step one can substitute
\beqn
t_1 &=& tx_1x_2x_3x_4, \\
t_2 &=& t(1-x_1)x_2x_3x_4, \non
t_3 &=& t(1-x_2)x_3x_4, \non
t_4 &=& t(1-x_3)x_4, \non
t_5 &=& t(1-x_4), \non
\int_0^\infty{\prod_{i=1}^5{(dt_i)} } &=&
\int_0^1{\prod_{i=1}^4{(dx_i)\ } x_2x_3^2x_4^3} \int_0^\infty{dt\ t^4} \nonumber
\eeqn
to do the trivial integration over the sum of all $t_i$, which gives 
\beqn \label{zwlointsub}
I_5^{(a)}(\eps) &=& \int_0^\infty{dt\ t^{-1-\eps}e^{-t}}
\int_0^1{\prod_{i=1}^4{(dx_i)}
  \frac{x_2x_3^2x_4^3 \bar{P}_5^{(a)}(x_1,x_2,x_3,x_4,\eps)}{\left( x_4
    (1-x_4(1-x_2x_3(1-x_2x_3))) \right) ^{5+\eps/2}} } \\ 
& & \times \left( \frac{\bar{P}_3(x_1,x_2,x_3,x_4)}{  x_4
        (1-x_4(1-x_2x_3(1-x_2x_3)))} \right)^\eps ,\nonumber
\eeqn
where the $\bar{P}^{(a)}$ are the numerators after the substitution. Investigating 
possible poles and finding them harmless we can expand the integrand and replace 
\beqn \label{subexpeins}
\left( \frac{\bar{P}_3(x_1,x_2,x_3,x_4)}{  x_4
        (1-x_4(1-x_2x_3(1-x_2x_3)))} \right)^\eps = 1+o(\eps) .
\eeqn
The integration over $t$ can be done using (\ref{gammafunktion}) and yields 
\beqn
\Gamma(-\eps) = -\frac{1}{\eps} -\gamma +o(\eps)  .
\eeqn
We shall later find that the first term of the integrand in (\ref{zwlointsub}) after 
substituting according to (\ref{subexpeins}) has only $1/\eps$ divergencies. This 
allows to extract the $1/\eps^2$ divergencies of the integrals, which originate 
from the poles of this first term, without regarding corrections proportional to 
$\eps$ from the second factor. This calculation is therefore able to reveal the 
leading order divergencies. It is completed in appendix \ref{fs}. There we also 
present an algorithm who enables an exact computation of all the integrals considered. 
Both methods lead to consistent results that are independently obtained. Hence 
we are very confident about the correctness of our calculations. \\ 

Regarding figure \ref{abbsewzwlogr} we notice that diagram (a) can have pinching 
limits conbtributing to diagram (c) if $t_i \rightarrow 0$ for $i\in \{ 1,...,4\}$, 
while it reduces to diagram (e) when $t_5 \rightarrow 0$. Further we have to 
perform the limit in a way that $t_1$ and $t_3 \rightarrow 0$ or $t_2$ and 
$t_4 \rightarrow 0$ to end up with diagram (d). The choices $t_1,t_2,t_3\rightarrow 0$ 
are obtained by assigning finite values to the sewing parameters $A_i$ and after 
permuting the integration variables they are all of the form 
\beqn
I^{(c)}_4(\eps) &=& \int{\frac{dA}{A^2(1-A)^2} \int_0^\infty{
    \prod_{i=1}^4{(dt_i)}
    \frac{P^{(c)}_4(t_1,...,t_4,A,\eps)}{\left(
      (t_1+t_2)(t_3+t_4)+t_3t_4 \right) ^{4+\eps/2} } }}\\
& & \times \exp \left( -\frac{P_3^{(c)}(t_1,...,t_4)}{
  (t_1+t_2)(t_3+t_4)+t_3t_4} \right). \nonumber
\eeqn
The exponent and the denominator are simply got by setting the appropriate proper 
time variables in (\ref{zwloaint}) to zero, whereas the numerator has to be 
computed separately by adding a power of $\al$ into the expansion of the integrand. 
The integration of those sewing variables that are left finite displays precisely 
the divergencies already encountered in the previous chapter when we discussed 
two-loop vacuum diagrams. We do not decide what to do with them at this stage but 
expand the integrand around $A=1$ and keep the finite as well as the divergent 
term. We then compute the proper time integration and mark the respective terms 
of the expansion in $A$ by their possibly divergent prefactors 
\beqn
O_2 \equiv \int{\frac{dA}{(1-A)^2}}, \quad O_0 \equiv \int{dA} .
\eeqn
We have for some cases even computed the integrations over proper times without doing 
any expansion of the integrand in the free moduli finding a more complicated 
structure of five instead of two 
terms but no deeper insight into the uncertainties of the pinching 
procedure. Therefore we restrict ourselves at this point to state the two types 
of terms we mentioned. The expression resulting from letting $t_4 \rightarrow 0$, which 
contributes to diagram (c), differs only in the respect, that keeping 
$k_2/(A_1A_2A_3)$ finite we end up with the undefined integral $dk_2/k_2^2$, 
which we, too, already found in the two-loop vacuum case. Then, of course, also 
$P_4^{(c)}$ depends on $k_2$ but is finite in the limit $k_2 \rightarrow 1$. 
We expand around $k_2=1$, mark the finite term by the factor 
\beqn
O_1 \equiv \int{dk_2}
\eeqn
and omit terms of the order $o(k_2-1)$. \\

We are now only left with contributions to (d) and (e). The case $t_5 \rightarrow 0$ 
is completely analogous to the one mentioned previously with the only exception 
that the integration looks somewhat easier:
\beqn
I^{(e)}_4(\eps) &=& \int{\frac{dk_1}{k_1^2} \int_0^\infty{
    \prod_{i=1}^4{(dt_i)}
    \frac{P^{(e)}_4(t_1,...,t_4,k_1,\eps)}{\left(
      (t_1+t_2)(t_3+t_4) \right) ^{4+\eps/2} } }} \exp \left(
-\frac{P_3^{(e)}(t_1,...,t_4)}{ (t_1+t_2)(t_3+t_4)} \right). 
\eeqn
We shall explore this by factorizing the integrals. In the last case of diagram (d) 
we do not find any contributions at all, simply because there are two proper time 
variables that are absent. This leads to two factors of $\al$ in front of the 
integral, which then vanishes proportional to the inverse string tension.\\
   
\begin{figure}[h]
\begin{picture}(400,70)

\LongArrow(50,30)(350,30)
\Line(50,30)(50,40)
\Line(100,30)(100,20)
\Line(150,30)(150,20)
\Line(200,30)(200,40)
\Line(250,30)(250,40)
\Vertex(370,30)1

\Text(40,48)[l]{$\eta_2=0$}
\Text(65,14)[l]{$z_2$}
\Text(91,14)[l]{$z_1$}
\Text(115,48)[l]{$\eta_1=\eta$}
\Text(140,48)[l]{$\xi_1=1$}
\Text(236,48)[l]{$\xi_2=\infty$}

\end{picture}
\caption[]{Parametrization (i) of world sheet (b)}  
\label{abbparambi}
\end{figure}
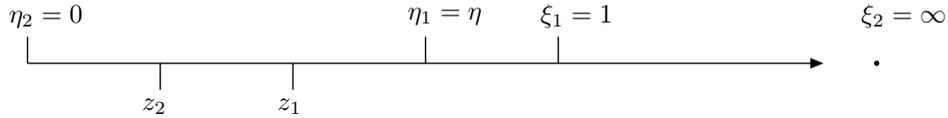

\begin{figure}[h]
\begin{picture}(400,70)

\LongArrow(50,30)(350,30)
\Line(50,30)(50,40)
\Line(100,30)(100,40)
\Line(150,30)(150,20)
\Line(200,30)(200,20)
\Line(250,30)(250,40)
\Vertex(370,30)1

\Text(40,48)[l]{$\eta_2=0$}
\Text(65,48)[l]{$\eta_1=\eta$}
\Text(90,14)[l]{$z_2$}
\Text(115,14)[l]{$z_1$}
\Text(140,48)[l]{$\xi_1=1$}
\Text(236,48)[l]{$\xi_2=\infty$}

\end{picture}
\caption[]{Parametrization (ii) of the world sheet (b)}  
\label{abbparambii}
\end{figure}
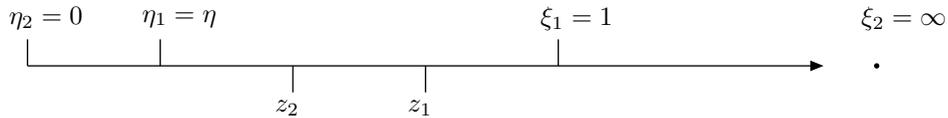

Having completed the study of contributions to diagram (a) and any diagrams 
including four-gluon vertices that arise from it, we now proceed to diagram (b) 
and its possible pinchings. We have to distinguish, which internal propagator 
the external states are sitting at, by regarding both of the two parametrizations 
drawn in figure \ref{abbparambi} and \ref{abbparambii}. \\

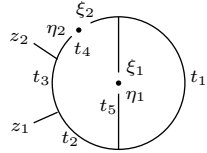
\begin{figure}[h]
\begin{picture}(400,70)

\CArc(200,30)(25,135,115)
\Line(168,45)(177,39)
\Line(168,15)(177,19)
\Line(200,5)(200,26)
\Line(200,34)(200,55)
\Vertex(200,30)1
\Vertex(185,50)1

\Text(200,25)[l]{\scriptsize{$\eta_1$}}
\Text(175,36)[l]{\scriptsize{$\xi_1$}}
\Text(106,13)[l]{\scriptsize{$z_1$}}
\Text(100,9)[l]{\scriptsize{$t_2$}}
\Text(69,50)[l]{\scriptsize{$\eta_2$}}
\Text(55,58)[l]{\scriptsize{$\xi_2$}}
\Text(73,32)[l]{\scriptsize{$t_1$}}
\Text(-12,32)[l]{\scriptsize{$t_3$}}
\Text(-46,47)[l]{\scriptsize{$z_2$}}
\Text(-37,21)[l]{\scriptsize{$t_5$}}
\Text(-72,43)[l]{\scriptsize{$t_4$}}

\end{picture}
\caption[]{Sewing (i) of diagram (b)}  
\label{abbsewgrbi}
\end{figure}  

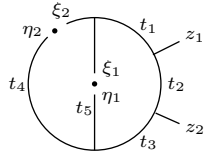
\begin{figure}[h]
\begin{picture}(400,70)

\CArc(200,30)(25,135,115)
\Line(232,46)(222,41)
\Line(223,19)(232,15)
\Line(200,5)(200,26)
\Line(200,34)(200,55)
\Vertex(200,30)1
\Vertex(185,50)1

\Text(200,25)[l]{\scriptsize{$\eta_1$}}
\Text(175,36)[l]{\scriptsize{$\xi_1$}}
\Text(164,8)[l]{\scriptsize{$t_3$}}
\Text(156,13)[l]{\scriptsize{$z_2$}}
\Text(69,50)[l]{\scriptsize{$\eta_2$}}
\Text(55,58)[l]{\scriptsize{$\xi_2$}}
\Text(63,52)[l]{\scriptsize{$t_1$}}
\Text(-12,30)[l]{\scriptsize{$t_4$}}
\Text(23,30)[l]{\scriptsize{$t_2$}}
\Text(-37,21)[l]{\scriptsize{$t_5$}}
\Text(-21,48)[l]{\scriptsize{$z_1$}}

\end{picture}
\caption[]{Sewing (ii) of diagram (b)}  
\label{abbsewgrbii}
\end{figure}   

Firgures \ref{abbsewgrbi} and \ref{abbsewgrbii} demonstrate how the left and right 
loop are to be distinguished. The sewing parameters of the configuration (i) are 
given by  
\beqn
\eta = A_1, \quad z_1 = A_1A_2, \quad z_2 = A_1A_2A_3,
\eeqn
those of (ii) by
\beqn
z_1 = A_1, \quad z_2 = A_1A_2, \quad \eta = A_1A_2A_3 .
\eeqn   
For both cases all but one proper time variables can be defined in a unique manner 
\beqn \label{sewb} 
t_i &=& -\al \ln (A_i) \quad \mbox{for}\ i=1,2,3, \\
t_4 &=& -\al \ln (k_2) -t_1-t_2-t_3 , \nonumber
\eeqn 
but the one of the smaller loop is different:
\beqn
t_5 &=& -\al \ln (k_1)-t_1 \quad \mbox{for (i)}, \\
t_5 &=& -\al \ln (k_1)-t_1-t_2-t_3 \quad \mbox{for (ii)} . \nonumber
\eeqn
This makes a distinction in the translation of further logarithms of moduli 
appearing in the integrand of the amplitude necessary. Configuration (i) has 
to be used with
\beqn
\al \ln (z_1-z_2) &=& -t_1-t_2+\al \ln (1-A_3) ,\\
\al \ln(z_1 z_2 ) &=& -2t_1-2t_2-t_3, \non
\al \ln \left( \frac{z_1}{z_2} \right) &=& t_3, \non
\al \ln \left( \frac{(z_1-\eta)(z_2-1)}{(z_1-1)(z_2-\eta)} \right) &=&
\al \ln \left( \frac{(1-A_2)(1-A_1A_2A_3)}{(1-A_1A_2)(1-A_2A_3)}
\right), \nonumber
\eeqn
whereas (ii) with:
\beqn
\al \ln (z_1-z_2) &=& -t_1+\al \ln (1-A_2) ,\\
\al \ln(z_1 z_2 ) &=& -2t_1-t_2, \non
\al \ln \left( \frac{z_1}{z_2} \right) &=& t_2, \non
\al \ln \left( \frac{(z_1-\eta)(z_2-1)}{(z_1-1)(z_2-\eta)} \right) &=&
 t_2+ \al \ln \left( \frac{(1-A_2A_3)(1-A_1A_2)}{(1-A_1)(1-A_3)}
\right). \nonumber
\eeqn
Despite the variables $t_2$ and $t_4$ featuring independently in the case (i), 
the integrand will only depend on their sum $t_2+t_4$, which is also 
true for $t_1$ and $t_3$ in the case of (ii). This reflects that the two 
propagators parametrized by these two proper times carry the same field theoretical 
momentum. By doing the usual expansion of the integrand we then obtain the amplitude 
\beqn \label{zwlobint}
{\cal A}^{(2)}_{2} \left( p^2 \right) \vert _{\mbox{\scriptsize (b)}} &=& N^2
\delta_{a_1 a_2} \frac{g^4}{(4\pi)^d} \int_0^\infty{\prod_{i=1}^5{(dt_i)} } \\
& & \times \frac{\left( (\eps_1p_2)(\eps_2p_1) P_5^{(b)}(t_1,...,t_5,d)
     +\eps^2  P_{4^\prime}^{(b)} (t_1,...,t_5,d)\right) }{\left(
    (t_1+t_5)(t_2+t_3+t_4)+t_1t_5 \right)^{d/2+3} } \non
& & \times\ \exp  \left( -p^2
\frac{t_3((t_1+t_5)(t_2+t_4)+t_1t_5)}{(t_1+t_5)(t_2+t_3+t_4)+t_1t_5}
\right) \non
&=&  N^2 \delta_{a_1 a_2}\left(  \frac{g}{4\pi} \right)^4 \left(
\frac{4\pi}{\sqrt{p^2}}\right)^{-\eps} \left( (\eps_1p_2)(\eps_2p_1)
I_5^{(b)}(\eps) + (\eps^2 p^2) I_{4^\prime}^{(b)}(\eps) \right) . \nonumber 
\eeqn
The sewing configurations (i) und (ii) lead to different explicit polynomials 
in the numerator but all  
are of identical structure, so that we do not have to treat them separately in 
this general discussion. We now follow the same strategies as earlier and cut 
our arguments short accordingly. From the figures \ref{abbsewgrbi} and 
\ref{abbsewgrbii} one immediately notices all relevant pinching contributions. 
Having $t_2 \rightarrow 0$ or $t_4\rightarrow 0$ in region (i) we get diagram (c), 
if both vanish (d). The same is true in region (ii) for $t_1$ and $t_3$. The 
moduli divergencies we find are identical to those found in the pinching 
contributions derived from diagram (a). We use again for vanishing $t_1$, 
$t_2$ or $t_3$ the prefactors $O_0$ and $O_2$ for the divergency and if $t_4$ 
is small we have $O_1$. A contribution to diagram (e) does not exist and 
contributions to (d) go to zero proportional to $\al$ again.\\

We have finally got all possible contributions to field theoretical Feynman diagrams 
that can be extracted from the appropriate string amplitude given a particular 
choice of local coordinates and fixed points. The results of the calculation of 
the integrals are summarized in table \ref{tabergeba} and
\ref{tabergebb}, while the details of the integration are postponed to appendix 
\ref{fs}. \\

\begin{table}[h]
\caption{Contributions to Feynman diagrams derived from (a)} \label{tabergeba}
\begin{center}
\begin{tabular}{llcc}
\vspace{0.2cm}
Diagram & Pinching & Order $\eps^{-2}$ &
Order $\eps^{-1}$ \\
\vspace{0.2cm}
(a) & & $\frac{35}{12}$ & $\frac{335}{24}$ \\
\vspace{0.2cm}
(c) & $t_1 \rightarrow 0$ & $ O_2 \frac{1}{6} + O_0 \frac{5}{16} $ 
& $O_2 \frac{5}{6} + O_0 \frac{37}{32} $ \\
\vspace{0.2cm}
(c) & $t_2 \rightarrow 0$ & $ O_2 \frac{1}{6} + O_0 \frac{5}{16} $ &
$ O_2 \frac{5}{6} + O_0 \frac{37}{32} $ \\
\vspace{0.2cm}
(c) & $t_3 \rightarrow 0$ & $  O_2 \frac{5}{24} + O_0 \frac{7}{16} $ 
& $ O_2 \frac{47}{48} + O_0 \frac{55}{32} $ \\  
\vspace{0.2cm}
(c) & $t_4 \rightarrow 0$ & $ O_1 \frac{25}{48} $ & $ O_1
\frac{205}{96} $ \\
\vspace{0.2cm}
(e) & $t_5 \rightarrow 0$ & $0$ & $0$ \\
\vspace{0.2cm}
(d) & $t_1,t_3 \rightarrow 0$ & $0$ & $0$ \\
\vspace{0.2cm}
(d) & $t_2,t_4 \rightarrow 0$ & $0$ & $0$ \\
\vspace{0.2cm}
\end{tabular}  
\end{center}
\end{table}

\begin{table}[h]
\caption{Contributions to Feynman diagrams derived from (b)} \label{tabergebb}
\begin{center}
\begin{tabular}{lllcc}
\vspace{0.2cm}
Diagram & Parametrization & Pinching & Order $\eps^{-2}$ &
Order $\eps^{-1}$ \\
\vspace{0.2cm}
(b) & (i) & & $ \frac{11}{6} $ & $ \frac{127}{18} $ \\
\vspace{0.2cm} 
(c) & (i) & $t_2 \rightarrow 0$ & $ O_2 \frac{1}{6} +O_0 \frac{5}{16}
$ & $ O_2 \frac{5}{6} +O_0 \frac{37}{32} $ \\
\vspace{0.2cm}
(c) & (i) & $t_4 \rightarrow 0$ & $O_1 \frac{7}{24} $ & $O_1
\frac{55}{48} $ \\
\vspace{0.2cm}
(d) & (i) & $t_2,t_4 \rightarrow 0$ & $0$ & $0$ \\
\vspace{0.2cm}
(b) & (ii) & & $0$ & $- \frac{14}{9} $ \\ 
\vspace{0.2cm}
(c) & (ii) & $t_1 \rightarrow 0$ & $ O_2 \frac{5}{24} +O_0
\frac{7}{16} $ & $ O_2 \frac{47}{48} + O_0 \frac{55}{32} $ \\
\vspace{0.2cm}
(c) & (ii) & $t_3 \rightarrow 0$ & $ O_2 \frac{5}{24} +O_0
\frac{7}{16} $ & $ O_2 \frac{47}{48} + O_0 \frac{55}{32} $ \\
(d) & (ii) & $t_1,t_3 \rightarrow 0$ & $0$ & $0$ \\
\end{tabular}  
\end{center}
\end{table}

All terms proportional to the Euler constant or $\ln \left( 4\pi/
p^2 \right) $ are dropped as they come out correctly automatically and can be 
restored easily. We have also changed our conventions in the favour of using 
$d=4-2\eps$ in order to compare results to the field theory. The given numbers 
still leave a lot of space to possible interpretations. Even only regarding 
the contributions to diagram (a) from figure \ref{abbgluonzweilo} reveal that 
there is no choice of the field theoretical gauge parameter, which lets the 
results coincide with the sum of the diagrams (i) + (j) + (k) of the same 
topology from table \ref{tabpeter}, if one also demands the sum of the two 
parametrizations (i) and (ii) of diagram (b) to coincide with (a) and (b) 
from table \ref{tabpeter}. An idenfication diagram by diagram seems to be 
ruled out therefore. The next would be to try to compare the sum of all 
diagrams. Unambiguously are
\beqn
\left( \frac{35}{12} + \frac{11}{6} \right) \eps^{-2} + \left(
  \frac{335}{24} + \frac{127}{18} - \frac{14}{9} \right) \eps^{-1} =
  \frac{ 19}{4} \eps^{-2} +\frac{467}{24} \eps^{-1} 
\eeqn
from diagram (a) and (b) without any four-gluon vertices, while the diagrams 
including such give 
\beqn \label{viererbeitrag}
\left( 3 \left( O_2 \frac{9}{24} + O_0 \frac{12}{16} \right) + O_1
  \frac{39}{48} \right) \eps^{-2} + \left( 3 \left( O_2 \frac{87}{48}
  + O_0 \frac{92}{32} \right) + O_1 \frac{315}{96} \right) \eps^{-1} .
\eeqn
They depend on the undefined diverging integrals over the free moduli, which 
should be replaced by some finite number in the manner of a 
regularization prescription. The most simple criterion for consistency could 
be seen to be the vanishing of the leading order of the expansion in $1/\eps$. 
As we deal with two unknown parameters this cannot give a unique answer and 
the general simplicity of the occurring numbers, which on the other hand 
might seem encouraging, allows more speculation about such a regularization 
prescription than we dare to present in this spot. For instance, using only the 
quadratic divergency $O_2$ and dropping $O_0$ completely, then demanding the $1/\eps^2$ 
term to vanish, one finds $(7/2) \eps^{-1}$. 
A surprisingly simple number, but it is simply wrong, 
neither is such a kind of discusion sufficiently rigorous in any manner.\\  

There are a couple of possibly sensible modifications of the general expansion 
method we used, that may have passed unnoticed in scalar theory and the 
one-loop Yang-Mills computation. One can for instance think of the ambiguity 
concerning which loop an external particle sits at and demand to add up all 
possibilities to attach a given number of external states to a diagram, by 
distinguishing all the different loops, even if the topolgy of the diagram 
created is identical. Following \cite{RS1} this would mean that one had to 
choose different local coordinates in the vicinity if the external states 
\beqn
V_i^\prime(0) = \left| \frac{(z_i-\eta_1)(z_i-\xi_1)}{\eta_1-\xi_1}
\right| = \left| \frac{(z_i-\eta)(z_i-1)}{\eta-1} \right| .
\eeqn
If we omit reducible diagrams, whose identification has been demonstrated in 
\cite{DV6}, the existing possibilities are listed diagrammatically in appendix 
\ref{sk}, where we also give the appropriate sewing translation and the results 
that are computed by solving the integrals which are found after expanding in 
all the relevant parameters. We have not displayed the pinching contributions 
which can be derived from these diagrams. The whole situation seems far too 
involved to allow any conclusion concerning the related field theoretical gauge 
choice and its relation to the coordinate fixing on the world sheet of the 
string, as the divergencies that are apearing when performing the pinching of 
these kind of diagrams are of the form
\beqn
\int{dA \frac{1}{A^3(1-A)}}  \quad \mbox{and}\ \int{dA
  \frac{1}{A(1-A)^3}} ,
\eeqn
which is different from that already encountered. We can neither draw any new 
conclusion with respect to the values of the ambiguous integrals.\\

\begin{center}
At the moment we are therefore unable to give any definite result.
\end{center}

\vspace{3cm}

\noindent {\bf Acknowledgements}
\vspace{.5cm}

We would like to thank Stefan Heusler for collaboration in earlier 
stages of this work and  
Markus Peter for substantial support in evaluating various integrations. M. G. Schmidt 
thanks Paolo Di Vecchia for exciting lectures and discussions 1996 in Heidelberg 
and kind hospitality in Copenhagen. We further thank him, A. Lerda, R. Marotta 
and R. Russo for important communications, as well 
as J. Fuchs and C. Schweigert for helpful explanations.

\clearpage

\begin{appendix} 
 
\section{The Schottky representation of Riemann surfaces}
\label{schottky}

\setcounter{equation}{0}
\renewcommand{\theequation}{A.\arabic{equation}}

The most important issues about the Schottky group and the representation of 
Riemann surfaces by its means are given in \cite{DV1,ALL}, while the method 
used to expand geometrical quantities in the Schottky multipliers is not 
published in the literature. We follow \cite{RO} in most respects. Some more 
general useful facts about abelian integrals and the prime form are detailed 
in \cite{MUM}.\\ 

\subsection{The Schottky group}

The convenient parametrization of automorphisms $t(z)$ of the compactified complex 
plane is written 
\beqn
t(z)= \frac{az+b}{cz+d} \quad {\mbox{with}} \quad a,b,c,d \in \mathbb{C} \quad
{\mbox{and}} \quad ad-bc=1. 
\eeqn
The matrices 
\beqn 
\left( \begin{array}{cc} 
a & b \\ c & d \end{array} \right) \quad \mbox{with} \quad \dt \left(
\begin{array}{cc} a & b \\ c & d \end{array} \right) =1    
\eeqn  
are in $SL(2,\mathbb{C})$, 
which is even isomorphic to the automorphism group. The Schottky 
pa\-ra\-me\-ters are then defined by 
\beqn
\frac{t(z)-\xi}{t(z)-\eta} = k\frac{z-\xi}{z-\eta}, 
\eeqn
where $\xi$ and $\eta$ are the fixed points of the mapping $t(z)$, $k$ its multiplier. 
The invariance under $\eta \leftrightarrow \xi$ and $k
\leftrightarrow k^{-1}$  allows to choose $\vert k \vert \leq 1$. The relation to the 
former parametrization is found to be 
\beqn \label{schottkyabb}
t(z) = \frac{\eta (z-\xi) -k \xi (z-\eta)}{(z-\xi)-k(z-\eta)}. 
\eeqn
And we read off:
\beqn
a &=& \frac{\eta-k\xi}{\sqrt{\vert k\vert} \vert \eta -\xi \vert}, \quad
b = \frac{-\eta \xi (1-k)}{\sqrt{\vert k \vert} \vert \eta -\xi \vert}, \non
c &=& \frac{1-k}{\sqrt{\vert k \vert} \vert \eta -\xi \vert}, \quad
d = \frac{k\eta-\xi}{\sqrt{\vert k \vert} \vert \eta -\xi
  \vert}. \nonumber 
\eeqn
The Schottky group $G_S^g$ is defined by composition and inversion of a given set of 
generators $\{ t_1,...,t_g\}$. It therefore consists of all mappings of the form 
$t^{j_1}_{i_1} t^{j_2}_{i_2} \cdots t^{j_m}_{i_m}$, having indices in $\{ 1,...,g\}$ 
and integer exponents.\\ 

Projective mappings transform circles into circles and for each there is a particular 
pair of isometric circles which have identical radii. The interior of the original 
circle is mapped onto the exterior of its image and the exterior of the origin 
onto the interior of its image. This induces a one to one identification of 
points and a fundamental region can be chosen to be the exterior of both circles  
which are identified. This construction of equivalence classes is a manifold that 
has the topology of a Riemann surface $\Sigma_g$ of Genus $g$, if this is the 
number of identified, disjoint circles. For technical reasons 
one excludes the set of points ${\cal D}$, where the orbits of the Schottky group become 
dense: 
\beqn
\{ C \cup \{\infty \} \} / G_S - {\cal D} \simeq \Sigma_g
\eeqn
The attractive and repulsive fixed points then lie inside the original, respectively 
the image isometric circle. Their radii $r$ and $\bar{r}$ and the coordinates of 
their centres are:
\beqn \label{isokr}
r &=& \bar{r} = \sqrt{\vert k \vert } \frac{\vert \eta -\xi \vert}{\vert
  1-k \vert}, \\
c &=& \frac{\xi -k\eta}{1-k}, \quad
\bar{c} = \frac{\eta-k\xi}{1-k}. \nonumber  
\eeqn

\begin{figure}
\begin{picture}(400,150)

\LongArrowArc(150,20)(20,270,269)
\LongArrowArc(250,60)(20,0,360)
\Vertex(150,20)1
\Vertex(250,60)1
\ArrowArcn(180,90)(75,322,262)

\DashCArc(200,40)(74,24,200)1
\DashCArc(200,40)(34,24,200)1

\Text(152,20)[l]{$c$}
\Text(226,60)[l]{$\bar{c}$}
\Text(116,3)[l]{$a_\mu$}
\Text(140,15)[l]{$b_\mu$}

\end{picture}
\caption[]{The non contractable cycles on a Torus} \label{abwege}
\end{figure}
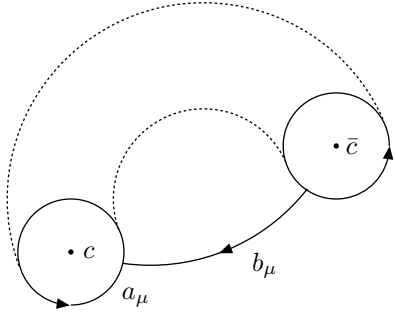

The dual of the homology basis of non contractable cycles is the cohomology 
basis in the sense of de Rham cohomology. 
It consists of the first abelian differentials 
\beqn
\omega_\mu \equiv \sum_{G_S(0,\mu)}{\left( \frac{1}{z-t(\eta_\mu)}
  -\frac{1}{z-t(\xi_\mu)} \right) } dz.
\eeqn
The sum is performed over all Schottky mappings that do not carry a power of $t_\mu$ 
on their right. The $\omega_\mu$ 
are analytic on all of $\Sigma_g$. One further has to demand the 
normalization of the integrals
\beqn
\oint_{a_\nu}{\omega^\mu}  = 2\pi i \delta_{\mu \nu}
\eeqn
along the paths $a_\nu$, that constitute one half of the homology basis to render the 
$\omega^\mu$ unique. The other half of the integrals 
\beqn
\oint_{b_\nu}{\omega_\mu} \equiv 2\pi i \tau_{\mu \nu}
\eeqn
are the entries of the period matrix $\tau_{\mu \nu}$ of the Riemann surface. Another 
geometrical function we shall need is the prime form which is characterized by its 
local behaviour  
\beqn
E(z_i,z_j) \rightarrow (z_i-z_j), \quad \mbox{for}\ z_i \rightarrow z_j
\eeqn 
and its transformation property under conformal changes of coordinates
\beqn
E\left( V_i(z_i),V_j(z_j) \right) = \frac{E(z_i,z_j)}{\sqrt{
    V^\prime_i \vert_{V_i^{-1}(z_i)} V^\prime_j \vert_{V_j^{-1}(z_j)}
    }} .
\eeqn
The defintion is unique, if also the invariance under transport around  $a_\nu$ 
cycles and a phase factor for $b_\nu$ cycles is demanded. One can always choose 
coordinates so that all the fixed points and the centre coordinates of isometrical 
circles are lying on the real axis, which greatly simplifies all figures we display. 
The coordinate invariance of string theory further allows to fix three of the moduli 
of the world sheet by global diffeomorphisms, which is conveniently employed to fix 
one at $0$, another at $\infty$ and finally one at $1$.\\ 

The two generalizations of the representation of Riemann surfaces one eventually 
needs in string theory are those to surfaces with boundaries and to surfaces that 
are not orientable. The former type of world sheets for open string theories is 
obtained by cutting holes into the complex plane, the closed string worldsheet. 
If one exclusively has to deal with open strings it is convenient to take only 
half of the complex plane and choose all the fixed points of the Schottky group 
that generates the desired loops as well as the insertion points of the external 
states to lie on its boundary, taken to be the real axis. Passing from points to 
equivalence classes then includes identifying $2g$ semi-circles whose centre 
coordinates are real. \\

\begin{figure}
\begin{picture}(400,120)

\LongArrow(60,10)(10,10)
\CArc(90,10)(30,360,180)
\Vertex(90,10)1
\Line(120,10)(150,10)
\CArc(160,10)(10,360,180)
\Vertex(160,10)1
\Line(170,10)(190,10)
\CArc(200,10)(10,360,180)
\Vertex(200,10)1
\Line(210,10)(250,10)
\CArc(280,10)(30,360,180)
\Vertex(280,10)1
\LongArrow(310,10)(380,10)
\DashCArc(185,10)(125,360,180)1
\DashCArc(185,10)(65,360,180)1
\DashCArc(180,10)(10,360,180)1
\DashCArc(180,10)(30,360,180)1

\Text(80,3)[l]{$\eta_2$}
\Text(126,3)[l]{$\eta_1$}
\Text(141,3)[l]{$\xi_1$}
\Text(195,3)[l]{$\xi_2$}

\end{picture}
\caption[]{The two-loop world sheet of an open string} \label{wsoffen}
\end{figure}
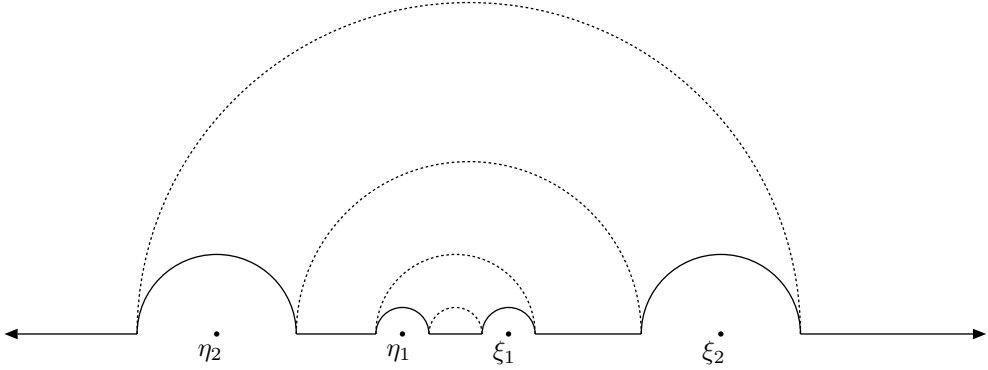

Non-orientable manifolds are constructed by including inversions into the generators 
of the Schottky group, which results in inserting projective planes 
$\mathbb{R}\mathbb{P}^2$ into the world sheet.

\subsection{Expanding in Schottky multipliers}
\label{schottkyentw}

We now demonstrate the techniques which allow to expand the geometrical quantities 
involved in the two-loop string amplitude into power series in the two Schottky 
multipliers $k_1$ and $k_2$, which then enables us to extract the relevant part 
of the amplitude. It is necessary to identify all terms up to the order 
$k_1^1k_2^1$, to cancel the prefactors coming from the integration measure 
of the amplitude.\\ 

We exploit the isomorphism from the projective group onto $SL(2,\mathbb{C})$ and use 
the basis invariance of trace and determinant. One can read off the matrix 
$T(z)$ corresponding to a Schottky generator $t(z)$ from (\ref{schottkyabb}) 
\beqn
T(z) = \frac{1}{\sqrt{\vert k \vert} \vert \eta -\xi \vert} \left(
\begin{array}{cc} 
\xi-k\eta & -\eta \xi (1-k) \\ 1-k & -\eta +k\xi  
\end{array} \right) 
\eeqn
and gets: 
\beqn
\frac{\dt \left( T(z) \right)}{(\tr \left( T(z)\right) )^2} =
\frac{k(\eta-\xi)^2}{(-(1+k)(\eta-\xi))^2} = \frac{k}{(1+k)^2} =
k+o(k^2). 
\eeqn
This gives a first order expression for the multiplier $k(t_1(z)t_2(z))$ 
of $t(z)=t_1(z)t_2(z)$:  
\beqn \label{multiplikator}
k(t_1(z)t_2(z)) &=& \frac{\dt (T_1(z)T_2(z))}{(\tr (T_1(z)T_2(z)))^2}
+o(k^2(t_1(z)),k^2(t_2(z))) \\ 
&=& \frac{\dt (T_1(z)) \dt (T_2(z))}{(\tr (T_1(z)T_2(z)))^2}
+o(k^2(t_1(z)),k^2(t_2(z)))  \non  
&=&
k(t_1(z))k(t_2(z))\frac{(\eta_1-\xi_1)^2(\eta_2-\xi_2)^2}
{(\xi_1-\eta_2)^2(\xi_2-\eta_1)^2}
+o(k^2(t_1(z)),k^2(t_2(z))), \nonumber 
\eeqn
where we had to use 
\beqn
\tr (T_1(z)T_2(z))= \frac{1}{\sqrt{\vert k_1 \vert \vert k_2 \vert }
  \vert \eta_1 -\xi_1 \vert \vert \eta_2 -\xi_2 \vert } \left( 
(\xi_1-\eta_2)(\xi_2-\eta_1)+o(k(t_1(z)),k(t_2(z))) \right) .
\eeqn
We simplify our notation and write $k_i \equiv k(t_i(z))$, dropping the argument 
$z$. Multipliers of mappings of the kind $t_1t_2^{-1}$ are obtained by replacing 
$\eta_2 \leftrightarrow \xi_2$ in (\ref{multiplikator}): 
\beqn \label{invmulti}
k(t_1t_2^{-1})= k_1k_2
\frac{(\eta_1-\xi_1)^2(\eta_2-\xi_2)^2}{(\xi_1-\xi_2)^2(\eta_1-\eta_2)^2} 
+o(k^2_1,k^2_2).
\eeqn
In fact, the two cases we have computed are already covering all types of multipliers 
needed. The symmetries of (\ref{multiplikator}) and (\ref{invmulti}) show that
\beqn
k(t_1t_2) &=& k(t_2t_1) =k(t_1^{-1}t_2^{-1}) =k(t_2^{-1}t_1^{-1}) ,\\ 
k(t_1^{-1}t_2) &=& k(t_1t_2^{-1}) =k(t_2^{-1}t_1) =k(t_2t_1^{-1})
\nonumber 
\eeqn
holds. One easily  deduces the complete set of Schottky mappings that might 
have multipliers which are only of first order in both the multipliers of the 
two generators to be the subset 
\beqn
G_S^2 \supset G_S^{(2)} \equiv \{ id, t_1, t_1^{-1}, t_2, t_2^{-1},
t_1t_2, t_2t_1, t_1^{-1}t_2, t_2t_1^{-1}, t_1t_2^{-1}, t_2^{-1}t_1 .
t_1^{-1}t_2^{-1}, t_2^{-1}t_1^{-1} \} \nonumber 
\eeqn
A useful asymptotic expression for the mapping itself is 
\beqn
t(z)=\xi +k(\eta-\xi)\frac{z-\xi}{z-\eta} +o(k^2),
\eeqn
which is obtained from (\ref{schottkyabb}). Because of $\vert k \vert \leq 1$ 
this explaines the asymptotic properties of a Schottky map:
\beqn
\lim_{n \rightarrow \infty} t^n(z) = \xi, \quad 
\lim_{n \rightarrow \infty} t^{-n}(z) = \eta.
\eeqn
We shall also need 
\beqn
\frac{(z_1-t(w_1))(z_2-t(w_2))}{(z_1-t(w_2))(z_2-t(w_1))}=
1+k\frac{(\xi-\eta)^2(z_1-z_2)(w_1-w_2)}{(z_1-\xi)(z_2-\xi)(w_1-\eta)(w_2-\eta)}
+o(k^2). \nonumber 
\eeqn
Now we can expand the period matrix, the abelian integrals, the normalization 
constant ${\cal N}$ of the partition function and the prime form to first order. 
An explicit expression for the prime form can easily be found from the definition 
given in the previous section:
\beqn  
E(z_1,z_2) \equiv (z_1-z_2) \prod_{t \in G_S  \setminus \{ id
  \} }{\sqrt{
    \frac{(z_1-t(z_2))(z_2-t(z_1))}{(z_1-t(z_1))(z_2-t(z_2))} }}, 
\eeqn
which has the expansion
\beqn
E^{(2)}(z_1,z_2) &=& (z_1-z_2) \prod_{t \in G^{(2)}_S \setminus \{ id
  \} }{\sqrt{
    \frac{(z_1-t(z_2))(z_2-t(z_1))}{(z_1-t(z_1))(z_2-t(z_2))} }}
+o(k_1^2,k_2^2) \\
&=& (z_1-z_2) \left( 1- k_1
\frac{(\eta_1-\xi_1)^2(z_1-z_2)^2}{(z_1-\eta_1)(z_2-\eta_1)(z_1-\xi_1)(z_2-\xi_1)}
\right) \non 
& & \times \left( 1- k_2
\frac{(\eta_2-\xi_2)^2(z_1-z_2)^2}{(z_1-\eta_2)(z_2-\eta_2)(z_1-\xi_2)(z_2-\xi_2)}
\right) \non 
& & \times \left( 1- k(t_1t_2)
\frac{(\eta_2-\xi_1)^2(z_1-z_2)^2}{(z_1-\eta_2)(z_2-\eta_2)(z_1-\xi_1)(z_2-\xi_1)}
\right) \non 
& & \times \left( 1- k(t_1t_2)
\frac{(\eta_1-\xi_2)^2(z_1-z_2)^2}{(z_1-\eta_1)(z_2-\eta_1)(z_1-\xi_2)(z_2-\xi_2)}
\right) \non 
& & \times \left( 1- k(t_1t_2^{-1})
\frac{(\xi_2-\xi_1)^2(z_1-z_2)^2}{(z_1-\xi_2)(z_2-\xi_2)(z_1-\xi_1)(z_2-\xi_1)}
\right) \non 
& & \times \left( 1- k(t_1t_2^{-1})
\frac{(\eta_2-\eta_1)^2(z_1-z_2)^2}{(z_1-\eta_2)(z_2-\eta_2)(z_1-\eta_1)(z_2-\eta_1)}
\right) +o(k_1^2,k_2^2). \nonumber 
\eeqn
The normalization constant of the partition function is being combined with two 
factors from the integration measure 
\beqn 
{\cal N} (1-k_1)^2 (1-k_2)^2 
\equiv \prod_{\beta}{\left(
  \prod_{n=1}^{\infty}{(1-k_\beta^n)^{-d}}
  \prod_{n=2}^{\infty}{(1-k_\beta^n)^2} \right)} (1-k_1)^2(1-k_2)^2, 
\eeqn
where the product over $\beta$ extends over the primary classes of the Schottky 
group, which are equivalence classes under conjugation. Those which contribute 
to the required order may be represented by
$G_S^{\mbox{\scriptsize pr}} \equiv \{ t_1, t_2, t_1t_2, 
t_1^{-1}t_2, t_1t_2^{-1} \}$. The expansion reads:
\beqn 
{\cal N}^{(2)} (1-k_1)^2 (1-k_2)^2 &=& \prod_{\beta \in G_S^{\mbox{\scriptsize pr}}}
{\left(
  \prod_{n=1}^{\infty}{(1-k_\beta^n)^{-d}}
  \prod_{n=2}^{\infty}{(1-k_\beta^n)^2} \right) } (1-k_1)^2(1-k_2)^2 +
o(k_1^2,k_2^2) \non
&=& 1+(d-2)(k_1+k_2) + \bigg( (d-2)^2+ d \bigg(
\frac{(\eta_1-\xi_1)^2(\eta_2-\xi_2)^2}{(\eta_1-\eta_2)^2(\xi_1-\xi_2)^2}
\non 
& &
+\frac{(\xi_1-\eta_1)^2(\xi_2-\eta_2)^2}{(\xi_1-\eta_2)^2(\xi_2-\eta_1)^2}
\bigg) \bigg) k_1k_2 + o(k_1^2,k_2^2).  
\eeqn
The integrals over the abelian differentials can be explicitely computed elementarily. 
They are then defined by sums over logarithms 
\beqn 
\int_{z_1}^{z_2}{\omega^\mu} \equiv \sum_{t \in G_S(0,\mu)}{ \ln
  \left(
    \frac{(z_2-t(\xi_\mu))(z_1-t(\eta_\mu))}{(z_1-t(\xi_\mu))(z_2-t(\eta_\mu))} 
\right) },  
\eeqn 
where the sum is performed over elements of $G^{(2)}_S$ which do not have a power of 
$t_\mu$ to the right, e.g. for $\mu=1$ just over $\{ id,
t_2, t_2^{-1}, t_1t_2, t_1^{-1}t_2, t_1t_2^{-1}, t_1^{-1}t_2^{-1}
\}$. The expansion is:
\beqn 
\exp \left( \int_{z_1}^{z_2}{\omega^1} \right)^{(2)} &=& \exp \left( \sum_{t
  \in G^{(2)}_S(0,1)}{ \ln \left(
  \frac{(z_2-t(\xi_1))(z_1-t(\eta_1))}{(z_1-t(\xi_1))(z_2-t(\eta_1))}
\right) } \right) +o(k_1^2,k_2^2) \\ 
&=& \frac{(z_2-\xi_1)(z_1-\eta_1)}{(z_2-\eta_1)(z_1-\xi_1)} \left( 1-
k_2
\frac{(\eta_2-\xi_2)^2(\xi_1-\eta_1)(z_1-z_2)}{(\eta_2-\xi_1)(\eta_2-\eta_1)
(z_1-\xi_2)(z_2-\xi_2)}
\right) \non 
& & \times \left( 1- k_2 
\frac{(\eta_2-\xi_2)^2(\xi_1-\eta_1)(z_1-z_2)}{(\xi_2-\xi_1)(\xi_2-\eta_1)
(z_1-\eta_2)(z_2-\eta_2)}
\right) \non 
& & \times \left( 1- k(t_1t_2) 
\frac{(\eta_2-\xi_1)^2(\xi_1-\eta_1)(z_1-z_2)}{(\eta_2-\xi_1)(\eta_2-\eta_1)
(z_1-\xi_1)(z_2-\xi_1)}
\right) \non 
& & \times \left( 1- k(t_1t_2)
\frac{(\xi_2-\eta_1)^2(\xi_1-\eta_1)(z_1-z_2)}{(\xi_2-\xi_1)(\xi_2-\eta_1)
(z_1-\eta_1)(z_2-\eta_1)}
\right) \non 
& & \times \left( 1- k(t_1t_2^{-1})
\frac{(\xi_2-\xi_1)^2(\xi_1-\eta_1)(z_1-z_2)}{(\xi_2-\xi_1)(\xi_2-\eta_1)
(z_1-\xi_1)(z_2-\xi_1)}
\right) \non 
& & \times \left( 1- k(t_1t_2^{-1})
\frac{(\eta_2-\eta_1)^2(\xi_1-\eta_1)(z_1-z_2)}{(\eta_2-\xi_1)(\eta_2-\eta_1)
(z_1-\eta_1)(z_2-\eta_1)}
\right) +o(k_1^2,k_2^2), \non 
\exp \left( \int_{z_1}^{z_2}{\omega^2} \right)^{(2)} &=&  \exp \left( \sum_{t
  \in G^{(2)}_S(0,2)}{ \ln \left(
  \frac{(z_2-t(\xi_2))(z_1-t(\eta_2))}{(z_1-t(\xi_2))(z_2-t(\eta_2))} 
\right) } \right) +o(k_1^2,k_2^2) \\   
&=& (1 \leftrightarrow 2). \nonumber 
\eeqn
The period matrix looks explicitly 
\beqn
2\pi i \tau_{\mu \nu} \equiv \delta_{\mu \nu} \ln (k_\mu) +\sum_{t \in
  G_S(\mu,\nu)}{\ln \left( 
  \frac{(\eta_\mu-t(\eta_\nu))(\xi_\mu-t(\xi_\nu))}{(\eta_\mu-t(\xi_\nu))
(\xi_\mu-t(\eta_\nu))} \right)}   
\eeqn
with summation over elements of $G^{(2)}_S$ that neither have a power of $t_\mu$ 
to the left nor one of $t_\nu$ to the right, for diagonal elements of 
$\tau_{\mu \nu}$ also $id$ is excluded. For instance for $\mu=1, \nu=1$ we 
only have to sum over $\{ t_2, t_2^{-1} \}$ while $\mu=1, \nu=2$ just allows 
$\{ id, t_2t_1, t_2t_1^{-1}, t_2^{-1}t_1, t_2^{-1}t_1^{-1} \}$. The expansion 
is obtained to be 
\beqn
2\pi i \tau_{11}^{(2)} &=& \ln (k_1) +\sum_{t \in G^{(2)}_S(1,1)}{\ln
  \left( 
  \frac{(\eta_1-t(\eta_1))(\xi_1-t(\xi_1))}{(\eta_1-t(\xi_1))(\xi_1-t(\eta_1))} 
\right)} +o(k_1^2,k_2^2) \non   
&=& \ln(k_1) + 2  
\frac{(\xi_1-\eta_1)^2(\xi_2-\eta_2)^2}{(\xi_1-\xi_2)(\eta_1-\eta_2)(\eta_1-\xi_2)
(\xi_1-\eta_2)}
+o(k_1^2,k_2^2), \non 
2\pi i \tau_{22}^{(2)} &=& (1 \leftrightarrow 2) , \non 
2\pi i \tau_{12}^{(2)} &=& 2\pi i \tau_{21}^{(2)} = \sum_{t \in
  G^{(2)}_S(1,2)}{\ln \left(
  \frac{(\eta_1-t(\eta_2))(\xi_1-t(\xi_2))}{(\eta_1-t(\xi_2))(\xi_1-t(\eta_2))} 
\right)} +o(k_1^2,k_2^2) \non 
&=& \ln \left(
\frac{(\eta_1-\eta_2)(\xi_1-\xi_2)}{(\eta_1-\xi_2)(\xi_1-\eta_2)}
\right) + 2 k(t_1t_2)
\frac{(\eta_1-\xi_1)(\xi_2-\eta_2)}{(\eta_1-\eta_2)(\xi_1-\xi_2)} \non
& & + 2 k(t_1t_2^{-1})
\frac{(\eta_1-\xi_1)(\xi_2-\eta_2)}{(\xi_1-\eta_2)(\eta_1-\xi_2)}
+o(k_1^2,k_2^2) .  
\eeqn
In all the given expressions we still have got the freedom to fix three of the four 
fixed points to arbitrary values and will conveniently choose these to be $0,1,\infty$.

\section{Computation of SPT integrals}
\label{fs}

\setcounter{equation}{0}
\renewcommand{\theequation}{B.\arabic{equation}}

In this appendix we explicitely compute the integrals over proper time variables 
we have obtained in chapter \ref{zwym} by performing the low energy limit of 
the two-loop string diagram extracting the gluon contribution. We use the 
conventional dimensional regularization scheme in $d=4+\eps$ dimensions. 
It is usually not necessary to distinguish between IR and UV by using different 
$\eps$ parameter as long as one can be sure that the overall amplitude is 
IR finite, which is obvious from (\ref{zwloaint}). For computing the integrals 
we therefore freely use the formulas (\ref{fsint1}), (\ref{fsint2}) and 
(\ref{fsint3}) by analytic extension outside the region, where the integrals 
on the left hand sides of these equations converge. We then finally expand 
the result in powers of $\eps$ and extract the minimal subtraction terms. 
This procedure is nothing but the usual one.

\subsection{Computation of the leading divergence}

We have found the term proportional to $(\eps p)(p\eps)$ in (\ref{zwloaint}) 
up to prefactors of the kind 
\beqn 
& & \int_0^\infty{ \prod_{i=1}^{5-v}{(dt_i)} \frac{\mbox{Polynomial of (5-$v$)th 
degree in} \ t_i} 
{(\mbox{Polynomial of second degree in} \ t_i)^{5-v+\eps/2}}} \\ 
& & \times \exp{\left( - \frac{\mbox{Polynomial of third degree in}
    \ t_i}{\mbox{Polynomial of second degree}\ t_i}\right) }, \nonumber 
\eeqn
if $v=0,1$ is the number of four gluon vertices we consider. The term proportional 
to $\eps^2p^2$ can be computed in completely the same manner and for the sake of 
brevity we concentrate on the one given above. If we are only interested in the 
leading $1/\eps^2$ divergence, we can meanwhile omitt the exponential from the 
integrand of (\ref{sptint}) and only after performing all but the last integration 
use it as an IR regulator. We 
have already done the substitution that lead to (\ref{zwlointsub}) and justified 
to replace the second term deriving from the exponent as in (\ref{subexpeins}) 
by $1+o(\eps)$. Then reverting the first substitution we get 
(\ref{zwloaint}) back up to the simplified exponent: 
\beqn \label{sptint}
\int_0^\infty{ \prod_{i=1}^{5-v}{(dt_i)} \frac{\mbox{Polynomial of (5-$v$)th 
      degree in} \ t_i} 
{(\mbox{Polynom of second degree in} \ t_i)^{5-v+\eps/2}}} \exp
\left( -\sum_{i=1}^5{t_i} \right). 
\eeqn
We actually need this integral only in an infinitesimal vicinity of the origin, 
where we are allowed to expand the integrand because of coninuity. If we take 
care of not getting any divergencies from large proper times we can finally 
extend the integration of the modified integrand to the full interval from zero 
to infinity again. We had to check that all the 
conditions mentioned are satisfied holding in 
all the cases we shall be regarding in the following. As the integrals obtained 
by these means can then be computed, we get a result for the $1/\eps^2$ divergent 
term that can be compared to results we shall get later on by more sophisticated 
methods that are less transparent on the other hand. 
We write the integrals thus obtained in the following manner
\beqn 
J_5(\eps) =\int_0^\infty{ \prod_{i=1}^5{(dt_i)} \frac{P_5(t_1,...,t_5,\eps)}
{(\poa)^{5+\eps
      /2}}}. 
\eeqn
In this subsection we use the notation $J^{(x)}_{5-v}\vert_{i...}(\eps)$ for 
these integrals without the exponential factor in the integrand, using the 
lower index for the number $5-v$ of integrations and the additional 
ones $i...$ for the possibly missing proper time variables, as well as an 
additional letter $(x)$ for the according diagram of figure \ref{abbgluonzweilo}, 
sticking strictly to the notation introduced in (\ref{sewa}) and (\ref{sewb}). 
In the end we shall reintroduce the exponential as cut-off without change of 
notation for the integral. The polynomials are carrying indices signaling their 
degree.\\ 

The diagram (a) has the contribution given in (\ref{zwloaint}) which has a polynomial  
\beqn \label{polnenner}
P^{(a)}_2(t_1,...,t_5) &=& \alq \left( \ln(k_1) \ln(k_2) -\ln^2(\eta) \right) \\
&=& (t_1+t_2)(t_3+t_4+t_5)+(t_3+t_4)t_5 \non 
&\equiv & a(t_1+t_2)+b \nonumber
\eeqn
in the denominator. For the polynomial $P^{(a)}_5$ in the numarator we use
\beqn
P^{(a)}_5(t_1,...,t_5,\eps) &=& \sum_{i=0}^3{c_i t_1^i} =
\sum_{i=0}^3{\sum_{j=0}^{3-i}{c_{ij} t_1^it_2^j}} =
\sum_{i=0}^3{\sum_{j=0}^{3-i}{\sum_{k=0}^{3,5-i-j}{c_{ijk}
      t_1^it_2^jt_3^k}}} \\ 
&=&
\sum_{i=0}^3{\sum_{j=0}^{3-i}{\sum_{k=0}^{3,5-i-j}{\sum_{l=0,2-i-j-k}^{3-k}
{c_{ijkl}t_1^it_2^jt_3^kt_4^l}}}}.
\nonumber 
\eeqn
The summation limits are to be understood in the sense that $0, 2-i-j-k$ stands 
for the larger value of $0$ or $2-i-j-k$, alternatively, if upper limits are 
double valued, they stand for the lower one. We now use for the integrations 
over $t_1$ and $t_2$ the formula (\ref{fsint1}), for that over $t_3$ we have 
(\ref{fsint2}) and finally for $t_4$ it is (\ref{fsint3}):  
\beqn 
J^{(a)}_5(\epsilon) &=& \int_0^\infty{\prod_{i=1}^5{(dt_i)} \frac{\sum_{i}{c_it_1^i}}
{(a(t_1+t_2)+b)^{5+\eps /2}}} \\
&=& \int_0^\infty{\prod_{i=2}^5{(dt_i)} \sum_{ij}{c_{ij}\ t_2^j\ (a t_2+
    b)^{i-4-\eps/2} a^{-i-1} B(i+1,4-i+\eps/2) }} \non 
&=& \int_0^\infty{dt_3dt_4dt_5 \sum_{ijk}{c_{ijk}\ t_3^k\
    b^{i+j-3-\eps/2} a^{-i-j-2} }} \non 
& &  \times B(i+1,4-i+\eps/2) B(j+1,3-i-j+\eps/2) \non
&=& \int_0^\infty{dt_4dt_5 \sum_{ijkl}{ c_{ijkl}\
    t_4^{i+j+k+l-2-\eps/2} t_5^{i+j-3-\eps/2} (t_4+t_5)^{-i-j-2} }}
\non 
& & \times B(i+1,4-i+\eps/2) B(j+1,3-i-j+\eps/2) B(k+1,4-k+\eps/2)
\non 
& & \times {_2}F_1(i+j+2,k+1;5+\eps/2;t_5/(t_4+t_5))  \non 
&=& \int_0^\infty{dt_5 \sum_{ijkl}{c_{ijkl}\ t_5^{i+j+k+l-6-\eps} 
    B(i+1,4-i+\eps/2) }} \non 
& & \times B(j+1,3-i-j+\eps/2) B(k+1,4-k+\eps/2) \non
& & \times B(3-k-l+\eps/2,i+j+k+l-1-\eps/2) \non
& & \times {_3}F_2(i+j+2,k+1,3-k-l+\eps/2;5+\eps/2,i+j+2;1). \nonumber 
\eeqn
For doing the $t_4$ integration we had to substitute
\beqn \label{substit}
t_4 &=& \frac{(1-x)t_5}{x}, \quad dt_4=-\frac{t_5dx}{x^2}, \quad
[0,\infty] \rightarrow [1,0]  .
\eeqn
The cut-off integral is  
\beqn \label{cutoff}
\int_{0}^\infty{dt_5\ t_5^{-1-\eps} e^{-t_5}} &=& \Gamma(-\eps) 
= -\frac{1}{\eps}- \gamma +o(\eps)\ . 
\eeqn 
To compute the integrals derived from (a) by pinching is even easier because of 
their polynomials being of lower degree. For $t_1 \rightarrow 0$ we get a 
contribution to diagram (c), having the polynomial   
\beqn
P^{(c)}_2(t_2,t_3,t_4,t_5) &=& t_2(t_3+t_4+t_5)+(t_3+t_4)t_5 \\
&\equiv & a t_2+b  \nonumber 
\eeqn
in the denominator and 
\beqn
P^{(c)}_4(t_2,t_3,t_4,t_5,\eps) &=& \sum_{i=0}^2{c_it_2^i} =
\sum_{i=0}^2{\sum_{j=0}^{1}{c_{ij}t_2^it_3^j}} =
\sum_{i=0}^2{\sum_{j=0}^{1}{\sum_{k=0,2-i-j}^{2,4-i-j}{c_{ijk}t_2^it_3^jt_4^k}}}
\eeqn
in the numerator, where all coefficients whose indices do not satisfy $j+k \leq 2$ 
vanish. The integrals are solved by 
(\ref{fsint1}),(\ref{fsint2}) and (\ref{fsint3}) using again the substitution 
(\ref{substit}). 
\beqn
J^{(c)}_4 \vert_1(\eps) &=& \int_{0}^\infty{\prod_{i=2}^5{(dt_i)}
  \frac{\sum_{i}{c_{i}t_2^i}}{(at_2+b)^{4+\eps/2}}} \\ 
&=&  \int_{0}^\infty{dt_3dt_4dt_5 \sum_{ij}{c_{ij}\ t_3^j\
    b^{i-3-\eps/2} a^{-i-1} B(i+1,3-i+\eps/2) }} \non 
&=&  \int_{0}^\infty{dt_4dt_5 \sum_{ijk}{c_{ijk}\ t_4^{i+j+k-2-\eps/2}
    t_5^{i-3-\eps/2} (t_4+t_5)^{-i-1} }} \non 
& & \times  B(i+1,3-i+\eps/2) B(j+1,3-j+\eps/2) \non
& & \times {_2}F_1(i+1,j+1;4+\eps/2;t_5/(t_4+t_5))  \non
&=&  \int_{0}^\infty{dt_5 \sum_{ijk}{c_{ijk}\ t_5^{i+j+k-5-\eps}
    B(i+1,3-i+\eps/2) B(j+1,3-j+\eps/2) }} \non 
& & \times B(i+j+k-1-\eps/2,2-j-k+\eps/2) \non
& & \times {_3}F_2(i+1,j+1,2-j-k+\eps/2;4+\eps/2,i+1;1) .\nonumber
\eeqn 
The $t_5$ integration again has to be treated by the cut-off prescription as in 
(\ref{cutoff}). The integral deduced from $t_2 \rightarrow 0$ has identical 
structure as this one with $t_1\rightarrow 0$, while $t_3\rightarrow 0$ can 
be handled even more simply. This then is equivalent to $t_4\rightarrow 0$ 
again. We get in both cases
\beqn 
P^{(c)}_2(t_1,t_2,t_4,t_5) &=& (t_1+t_2)(t_4+t_5)+t_4t_5 \\
&\equiv & a(t_1+t_2)+b \nonumber 
\eeqn
in the denominator and 
\beqn
P^{(c)}_4(t_1,t_2,t_4,t_5,\eps) &=& \sum_{i=0}^1{c_i t_1^i} =
\sum_{i=0}^1{\sum_{j=0}^{2-i}{ c_{ij} t_1^i t_2^j}} =
\sum_{i=0}^1{\sum_{j=0}^{2-i}{\sum_{k=2-i-j}^{2,4-i-j}{c_{ijk} t_1^i
      t_2^j t_4^k}}} 
\eeqn
in the numerator. The first integrations can all be done applying (\ref{fsint1}). 
Afterwards the cut-off is employed and the result reads
\beqn
J^{(c)}_4 \vert_3(\eps) &=& \int_0^\infty{dt_1dt_2dt_4dt_5 \frac{\sum_{i}{c_i
      t_1^i}}{(a(t_1+t_2) +b)^{4+\eps/2}}} \\ 
&=&  \int_0^\infty{dt_2dt_4dt_5 \sum_{ij}{c_{ij}\ t_2^j\ a^{-i-1}
    (at_2+b)^{i-3-\eps/2} B(i+1,3-i+\eps/2) }} \non 
&=&  \int_0^\infty{dt_4dt_5 \sum_{ijk}{c_{ijk}\ t_4^{i+j+k-2-\eps/2}
    t_5^{i+j-2-\eps/2} (t_4+t_5)^{-i-j-2} }} \non 
& & \times B(i+1,3-i+\eps/2) B(j+1,2-i-j+\eps/2) \non
&=&  \int_0^\infty{dt_5 \sum_{ijk}{c_{ijk}\ t_5^{i+j+k-5-\eps}
    B(i+1,3-i+\eps/2) }} \non 
& & \times B(j+1,2-i-j+\eps/2) B(i+j+k-1-\eps/2,3-k+\eps/2) .\nonumber
\eeqn
We shall not present the computation 
of the contributions to diagram (e) as they are vanishing, 
which makes it a little boring. This is in fact a consequence 
of numerous cancellations, whereas the vanishing of contributions to diagram 
(d) is due to the overall prefactor $\al$ as already mentioned.\\ 

We now proceed to diagram (b) of figure \ref{abbgluonzweilo}, where we get 
\beqn
P_2^{(a)} (t_1,...,t_5) &=& \alq \left( \ln(k_1) \ln(k_2)
-\ln^2(\eta) \right) \\ &=& (t_1+t_5)(t_2+t_3+t_4)+t_1t_5 \equiv
a(t_2+t_3+t_4) +b \nonumber  
\eeqn
in the denominator which suggests to integrate by the order $t_2 \rightarrow 
t_3 \rightarrow t_4 \rightarrow t_1$. The polynomial in the numerator then is    
\beqn
P_5^{(b)} (t_1,...,t_5,\eps)   &=& \sum_{i=0}^3{c_i t_2^i} =
\sum_{i=0}^3{\sum_{j=0}^{3-i}{c_{ij} t_2^it_3^j}} \\ 
&=& \sum_{i=0}^3{\sum_{j=0}^{3-i}{\sum_{k=0}^{3-i-j}{c_{ijk} 
      t_2^it_3^jt_4^k}}} =
\sum_{i=0}^3{\sum_{j=0}^{3-i}{\sum_{k=0}^{3-i-j}{\sum_{l=0,2-i-j-k}^{3,5-i-j-k}
{c_{ijkl}t_2^it_3^jt_4^kt_1^l}}}} .\nonumber  
\eeqn
The integrations can be performed be using exclusively (\ref{fsint1}) and reveal 
\beqn 
\bar{J}^{(b)}_5(\eps) &=& \int_0^\infty{dt_1dt_2dt_3dt_4dt_5
  \frac{\sum_{i}{c_it_2^i}}{(a(t_2+t_3+t_4)+b)^{5+\eps/2}}} \\ 
&=&   \int_0^\infty{dt_1dt_3dt_4dt_5 \sum_{ij}{c_{ij}\ t_3^j\ a^{-i-1}
    (a(t_3+t_4)+b)^{i-4-\eps/2} }} \non 
& & \times B(i+1,4-i+\eps/2) \non
&=&  \int_0^\infty{dt_1dt_4dt_5 \sum_{ijk}{c_{ijk}\ t_4^k\ a^{-i-j-2}
    (at_4+b)^{-i-j-k-3} }} \non 
& & \times B(i+1,4-i+\eps/2) B(j+1,3-i-j+\eps/2) \non 
&=&  \int_0^\infty{dt_1dt_5 \sum_{ijkl}{c_{ijkl}\
    t_1^{i+j+k+l-2-\eps/2} t_5^{i+j+k-2-\eps/2} (t_1+t_5)^{-i-j-k-3}
    }} \non  
& & \times B(i+1,4-i+\eps/2) B(j+1,3-i-j+\eps/2) \non 
& & \times B(k+1,2-i-j-k+\eps/2) \non 
&=&  \int_0^\infty{dt_5 \sum_{ijkl}{c_{ijkl}\ t_5^{i+j+k+l-6-\eps}
    B(i+1,4-i+\eps/2) }} \non  
& & \times B(j+1,3-i-j+\eps/2) B(k+1,2-i-j-k+\eps/2) \non 
& & \times B(4-l+\eps/2,i+j+k+l-1-\eps/2) ,\nonumber 
\eeqn 
where again the final $t_5$ integration has to be done by (\ref{cutoff}). 
All contributions to diagrams including four gluon vertices can be reduced 
to cases we have already covered when discussing those derived from (a). 
The necessary replacements are quite obvious.

\subsection{Exact computation of SPT integrals}

For the integrals from (\ref{zwloaint}) we can define an algorithm that even allows 
an exact computation. This will reveal further information to us, which hints, 
how the rest of the integrals corresponding to all the other diagrams can be 
solved by explicit calculation, using only the formulas displayed in appendix 
\ref{formeln}. We have to thank M. Peter from Heidelberg for his advice 
and active help on this subject. Our notation is adopted from the previous chapter, only 
replacing $J$ by $I$. \\

The most important observation is that the proper time integrals can formally 
be written as ordinary momentum integrals of a scalar field theory, which are 
known from solving the integrals emerging from Feynman rules for such a theory. 
To do this we have to treat each term of the polynomials in the numerators 
separately and exploit the identity 
\beqn
I(q^2,n_1,...,n_5) &=& \int_{-\infty}^\infty{\frac{d^Dp
    d^Dk}{(2\pi)^{2D}} \frac{1}{\left( p^2 \right)^{n_1} \left(
    (p-q)^2 \right) ^{n_2} \left( k^2 \right) ^{n_3} \left( (k-q)^2
  \right) ^{n_4} \left( (p-k)^2 \right) ^{n_5} } } \\
&=& \int_0^\infty{\prod_{i=1}^5{\left( \frac{dt_i\ t_i^{n_i}}{\Gamma (n_i) }
  \right) } \left( (t_1+t_2)(t_3+t_4+t_5)+(t_3+t_4)t_5 \right) ^{-D/2}} \non
& & \times\ \exp \left( -q^2 \frac{t_1t_2(t_3+t_4+t_5)+t_1t_3(t_4+t_5)+
t_2t_4(t_3+t_5)+t_3t_4t_5}{(t_1+t_2)(t_3+t_4+t_5)+(t_3+t_4)t_5}
\right) .  \nonumber
\eeqn
The SPT integral of the second line with $D=10+\eps$ exactly resembles the 
integrals we are trying to solve when dealing with Yang-Mills theory. We use 
the variable $t_i$ for the propagator with exponent $n_i$ as in (\ref{impspt}) 
and then do the Gaussian integrations by (\ref{gaussint}). Such momentum 
integrals one knows how to handle. The recursion relation
\beqn
0 &=& \int_{-\infty}^\infty{\frac{d^Dp d^Dk}{(2\pi)^{2D}}
  \frac{\partial}{\partial p_\mu}  \left( \frac{ p_\mu-k_\mu  }{\left(
  p^2 \right)^{n_1} \left( (p-q)^2 \right) ^{n_2} \left( k^2 \right)
  ^{n_3} \left( (k-q)^2 \right) ^{n_4} \left( (p-k)^2 \right) ^{n_5} }
\right) } \\
&=& (D-n_1-n_2-2n_5) I(q^2,n_1,...,n_5) \non
& & -n_1 \left( I(q^2,n_1+1,n_2,n_3,n_4,n_5-1) - I(q^2,n_1+1,n_2,n_3-1,n_4,n_5)
\right) \non
& & - n_2 \left( I(q^2,n_1,n_2+1,n_3,n_4,n_5-1) -
I(q^2,n_1,n_2+1,n_3,n_4-1,n_5) \right) \nonumber 
\eeqn
relates any given integral successively to a number of integrals that has $n_5=0$, 
$n_3=0$ or $n_4=0$. In the first case both integrals factorize as one propagator 
vanishes, in the latter case we get two one-loop integrals that can be easily 
computed, which will be done explicitly in the following. 
This recursion thus allows to reduce all terms to a small number of 
simple types, but it creates a couple of thousand terms of such. 
Therefore we have used the algebraic computer programs MAPLE and FORM to do the task 
of performing the recursion and the expansion of its results in $1/\eps$. The 
results are summarized in several tables chapter \ref{zwym}.\\

We next demonstrate how the pinching contributions of (a) can be calculated and 
shall find that we have to explicitely compute only the two intgrals just cited. 
To find the most appropriate order of substitutions for the proper time variables 
it is essential to understand the translation into momentum integrals. Take the 
case $t_1 \rightarrow 0$ whose contribution is 
\beqn \label{intic1}
I^{(c)}_4\vert_1 (\eps) &=& \int_0^\infty{\prod_{i=2}^5{\left( dt_i
  \right) } \frac{ P_4^{(c)}(t_2,...,t_5,\eps) }{\left(
  t_2(t_3+t_4+t_5)+(t_3+t_4)t_5 \right)^{4+\eps/2} }} \\
& & \times\ \exp \left( -
\frac{t_2t_4(t_3+t_5)+t_3t_4t_5}{t_2(t_3+t_4+t_5)+(t_3+t_4)t_5}
\right) \nonumber
\eeqn
and use 
\beqn
t_2 = xt, \quad t_5 = (1-x)t, \quad t_3 \rightarrow tt_3, \quad t_4
\rightarrow tt_4 ,
\eeqn 
rescaling $t_2$ and $t_5$ to a unit square. On the contrary, a simultaneous scaling 
of all the integration variables would have not been successfull. The integration 
over $t$ is trivial and gives 
\beqn
I^{(c)}_4\vert_1 (\eps) = \Gamma (-\eps) \int_0^1{dx
  \int_0^\infty{dt_3dt_4\ \left( \frac{t_3t_4 + x(1-x)t_4
      }{t_3+t_4+x(1-x) } \right)^\eps
    \frac{\bar{P}^{(c)}_4(t_3,t_4,x,\eps) }{\left( t_3+t_4+x(1-x)
    \right) ^{4+\eps/2}}}} .
\eeqn
We next define  
\beqn
\bar{P}_4^{(c)}(t_3,t_4,x,\eps) \equiv \sum_i{c_i (t_3,x,\eps) t_4^i}
  \equiv \sum_{ij}{c_{ij}(x,\eps) t_4^i t_3^j} \equiv
    \sum_{ijk}{c_{ijk}(\eps) t_4^i t_3^j x^k} 
\eeqn
and integrate over $t_3$ and $t_4$ by using (\ref{fsint1}):
\beqn
I^{(c)}_4\vert_1 (\eps) &=& \Gamma (-\eps) \int_0^1{dx\
  \sum_{ijk}{c_{ijk}(\eps) x^{i+j+k-2+\eps/2 } (1-x)^{i+j-2+\eps/2} }} \\
& & \times\ B\left( i+1+\eps, 3-i+\eps/2 \right) B\left( j+1,2-i-j-\eps/2
  \right)  \non
&=&  \Gamma (-\eps) \sum_{ijk}{c_{ijk}(\eps)  B\left( i+1+\eps,
  3-i+\eps/2 \right) B\left( j+1,2-i-j-\eps/2 \right) } \non
& & \times\ B\left( i+j-1+\eps/2, i+j+k-1+\eps/2 \right) . \nonumber
\eeqn
The cases $t_2,t_3,t_4 \rightarrow 0$ can be treated in precisely the same 
manner by a simple permutation of indices, which can be read of from (\ref{zwloaint}) 
by inspecting (\ref{intic1}). We only remain with computing the contribution 
coming from $t_5 \rightarrow 0$ to the diagram (e):
\beqn
I^{(e)}_4\vert_5 (\eps) &=& \int_0^\infty{\prod_{i=1}^4{\left( dt_i
  \right) } \frac{ P_4^{(e)}(t_1,...,t_4,\eps) }{\left(
  (t_1+t_2)(t_3+t_4) \right)^{4+\eps/2} }} \\
& & \times\ \exp \left( -
\frac{  t_1(t_2t_3+t_2t_4+t_3t_4)+t_2t_3t_4 }{(t_1+t_2)(t_3+t_4)}
\right) \non
&=& \int_0^\infty{\prod_{i=1}^4{\left( dt_i \right) } \frac{
    P_4^{(e)}(t_1,...,t_4,\eps) }{\left( (t_1+t_2)(t_3+t_4)
  \right)^{4+\eps/2} }} \exp \left( - \frac{t_1t_2}{t_1+t_2} +
\frac{t_3t_4}{t_3+t_4} \right) .\nonumber 
\eeqn
We rescale $t_1$ and $t_2$  
\beqn
t_1 = xt, \quad t_2 = (1-x)t ,
\eeqn 
and as the polynomial $P^{(e)}_4$
in the numerator is always quadratic in these two variables, 
the integration over $t$ can be split off from the rest:
\beqn
I^{(e)}_4\vert_5 (\eps) &=& \Gamma \left( -\eps/2 \right) \int_0^1{dx
  \int_0^\infty{dt_3 dt_4\ \left( x(1-x) \right) ^{\eps/2} }} \\
& & \times\ \frac{\bar{P}^{(e)}_4(t_3,t_4,x,\eps)}{(t_3+t_4)^{4+\eps/2}} \exp
    \left( -\frac{t_3 t_4}{t_3+t_4} \right) .\nonumber
\eeqn
After another substitution
\beqn
t_3 = y\rho, \quad t_4 = (1-y)\rho ,
\eeqn 
and defining
\beqn
\bar{P}^{(e)}_4(t_3,t_4,x,\eps) \equiv \sum_i{c_i(y,\eps) x^i} \equiv
\sum_{ij}{c_{ij}(\eps) x^i y^j }
\eeqn
we finally obtain
\beqn
I^{(e)}_4\vert_5 (\eps) &=& \Gamma \left( -\eps/2 \right)^2 \int_0^1{dxdy\ 
\sum_{ij}{c_{ij}(\eps) \left( x(1-x) \right) ^{\eps/2}
    \left( y(1-y) \right) ^{\eps/2} x^i y^j  }} \\
&=&  \Gamma \left( - \eps/2 \right)^2 \sum_{ij}{c_{ij} (\eps) B\left(
  i+1+\eps/2,1+\eps/2 \right) B\left( j+1+\eps/2, 1+\eps/2 \right) }. \nonumber
\eeqn
We have such got all the pinching contributions derived from (a).\\

The only case we have not covered yet is the diagram (b) itself. All its 
pinching limits can again be reduced to the former $t_1 \rightarrow 0$ 
integral type. From (\ref{zwlobint}) we get, following the notation of 
figure \ref{abbsewgrbi}, 
\beqn
I_5^{(b)}(\eps) &=& \int_0^\infty{\prod_{i=1}^5{(dt_i)} \frac{P_5^{(b)}
(t_1,...,t_5,\eps)}{\left( (t_1+t_5)(t_2+t_3+t_4)+t_1t_5 \right)^{5+\eps/2} }} \\
& & \times\ \exp  \left( -
\frac{t_3((t_1+t_5)(t_2+t_4)+t_1t_5)}{(t_1+t_5)(t_2+t_3+t_4)+t_1t_5}
\right) . \nonumber
\eeqn
In fact, by explicit inspection the polynomial $P_5^{(b)}$ turns out to 
depend only on the sum of $t_2$ and $t_4$. This corresponds to the fact 
that from the point of view of the field theory, one of the two variables 
is superfluous, as both propagators carry the same momentum, are thus 
identical. We then use their sum as a new variable 
\beqn
t_2 = xt, \quad t_4 = (1-x)t , \quad t_1 \rightarrow tt_1 , \quad t_3
\rightarrow tt_3 , \quad t_5 \rightarrow tt_5 .
\eeqn 
The integrand does not depend on $x$ and the integration over $t$ is trivial, 
so that we get
\beqn
I_5^{(b)}(\eps) &=& \Gamma (-\eps) \int_0^\infty{ dt_1dt_3dt_5\
  \frac{\bar{P}_5^{(b)}(t_1,t_3,t_5,\eps)}{\left(
    (t_1+t_5)(t_3+1)+t_1t_5 \right) ^{5+\eps/2} }} \\
& & \times\ \left( \frac{t_3(t_1+t_5+t_1t_5)}{
    (t_1+t_5)(t_3+1)+t_1t_5} \right) ^\eps . \nonumber
\eeqn
We then proceed as usual
\beqn
\bar{P}_5^{(b)}(t_1,t_3,t_5,\eps) \equiv \sum_i{c_i(t_1,t_5,\eps)
  t_3^i} \equiv \sum_{ij}{c_{ij}(t_5,\eps) t_3^i t_1^j} \equiv
\sum_{ijk}{c_{ijk} (\eps) t_3^i t_1^j t_5^k} ,
\eeqn
integrate over $t_3$ by (\ref{fsint1}), next over $t_1$ using (\ref{fsint2}) 
and finally over $t_5$ substituting 
\beqn
t_5 = \frac{y}{1-y} , \quad dt_5 = \frac{dy}{(1-y)^2} , \quad
[0,\infty] \rightarrow [0,1] 
\eeqn 
and by (\ref{fsint3}). The result of all this is
\beqn
I_5^{(b)}(\eps) &=& \Gamma (-\eps) \sum_{ijk}{c_{ijk}(\eps) B\left(
  i+1+\eps, 4-i+\eps/2 \right) B\left( j+1, 4-j+\eps/2 \right) } \\
& & \times\ B\left( j+k-3-\eps/2,4-k+\eps/2 \right) \non
& & \times\ {_{3}}F_{2} \left( i+1+\eps, j+1, j+k-3-\eps/2; 5+\eps/2,
j+1; 1\right) . \nonumber
\eeqn
We have been able to exactly solve all the SPT integrals.

\section{Two-loop diagrams with external states attached to the small loop} 
\label{sk}

\setcounter{equation}{0}
\renewcommand{\theequation}{C.\arabic{equation}}

This is an extension to chapter \ref{zwym}, where we treat diagrams with one or 
both external states sitting at the interior, ``small'' loop. The procedure is very 
much equivalent to the usual and differs mainly in the choice of local coordinates 
one uses in the vicinity of the external legs of the diagrams, as well as in the 
parametrization of the sewing variables that is derived fom this. We shall 
therefore be very brief and present all possible combinations of contributions 
in terms of the world sheet diagrams we have introduced earlier, then display 
in table \ref{tabparazwloanh} the appropriate sewing parameters and finally 
simply cite the results one obtains from solving the integrals in table 
\ref{tabbeitragzwloanh}. For the local coordinates of those external states 
that are supposed to be attached to the interior loop we now take 
\beqn
V_i^\prime(0) = \left| \frac{(z_i-1)(z_i-\eta)}{1-\eta} \right| ,
\eeqn
while for those which are sitting at the large loop we have $V_i^\prime(0) =z_i$ 
as before. The parametrization we define in table \ref{tabparazwloanh} is given 
by the methods of \cite{RS1}, which necessarily leads to the correct \gf.

\begin{figure}[h]
\begin{picture}(400,140)

\LongArrow(50,30)(350,30)
\Line(50,30)(50,40)
\Line(250,30)(250,90)
\Line(120,30)(120,40)
\Line(180,20)(180,30)
\Line(240,60)(250,60)
\Vertex(370,30)1

\Text(40,48)[l]{$\eta_2=0$}
\Text(85,48)[l]{$\eta_1=\eta$}
\Text(121,14)[l]{$z_1$}
\Text(150,60)[l]{$z_2$}
\Text(140,98)[l]{$\xi_1=1$}
\Text(236,48)[l]{$\xi_2=\infty$}

\end{picture}
\caption[]{Parametrization (I) of the world sheet of diagram (a)}  
\label{abbparamaII}
\end{figure}
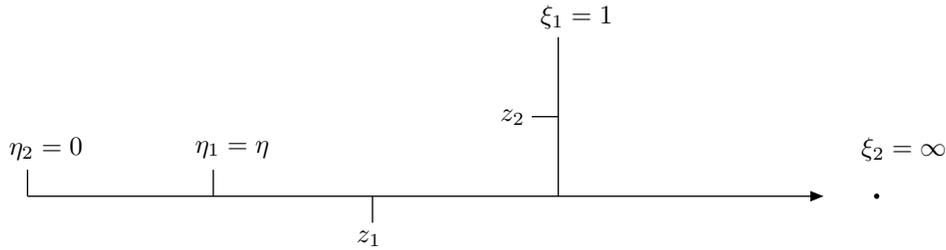     

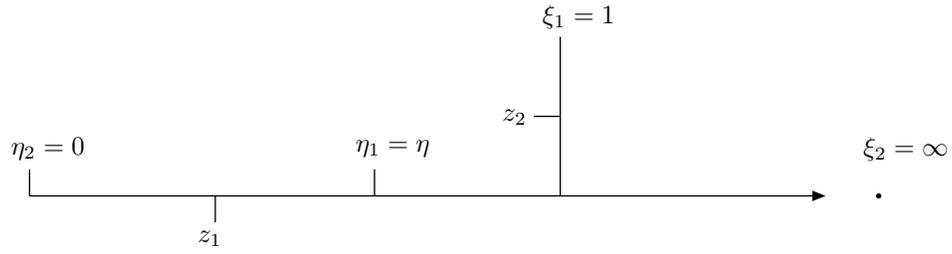
\begin{figure}[h]
\begin{picture}(400,140)

\LongArrow(50,30)(350,30)
\Line(50,30)(50,40)
\Line(250,30)(250,90)
\Line(120,20)(120,30)
\Line(180,40)(180,30)
\Line(240,60)(250,60)
\Vertex(370,30)1

\Text(40,48)[l]{$\eta_2=0$}
\Text(145,48)[l]{$\eta_1=\eta$}
\Text(60,14)[l]{$z_1$}
\Text(150,60)[l]{$z_2$}
\Text(140,98)[l]{$\xi_1=1$}
\Text(236,48)[l]{$\xi_2=\infty$}

\end{picture}
\caption[]{Parametrization (II) of the world sheet of diagram (a)}  
\label{abbparamaIII}
\end{figure}     
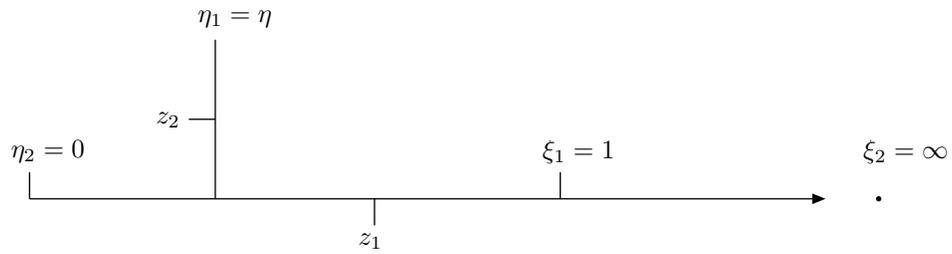
\begin{figure}
\begin{picture}(400,140)

\LongArrow(50,30)(350,30)
\Line(50,30)(50,40)
\Line(250,30)(250,40)
\Line(120,30)(120,90)
\Line(180,20)(180,30)
\Line(110,60)(120,60)
\Vertex(370,30)1

\Text(40,48)[l]{$\eta_2=0$}
\Text(146,14)[l]{$z_1$}
\Text(60,98)[l]{$\eta_1=\eta$}
\Text(19,60)[l]{$z_2$}
\Text(140,48)[l]{$\xi_1=1$}
\Text(236,48)[l]{$\xi_2=\infty$}

\end{picture}
\caption[]{Parametrization (III) of the world sheet of diagram (a)}  
\label{abbparamaIV}
\end{figure}     
\begin{figure}[h]
\begin{picture}(400,140)

\LongArrow(50,30)(350,30)
\Line(50,30)(50,40)
\Line(250,30)(250,40)
\Line(120,30)(120,20)
\Line(180,90)(180,30)
\Line(180,60)(170,60)
\Vertex(370,30)1

\Text(40,48)[l]{$\eta_2=0$}
\Text(86,14)[l]{$z_1$}
\Text(120,98)[l]{$\eta_1=\eta$}
\Text(79,60)[l]{$z_2$}
\Text(140,48)[l]{$\xi_1=1$}
\Text(236,48)[l]{$\xi_2=\infty$}

\end{picture}
\caption[]{Parametrization (IV) of the world sheet of diagram (a)}  
\label{abbparamaV}
\end{figure}
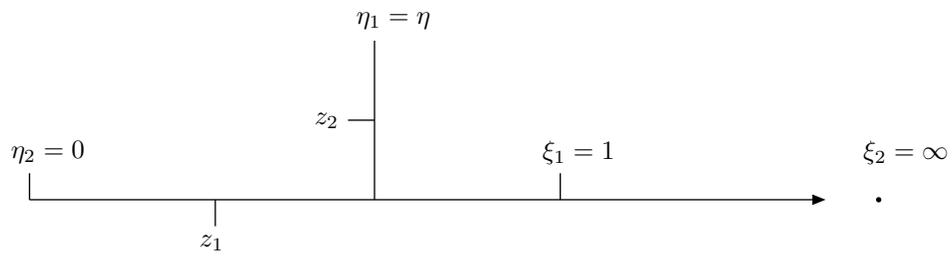     

\begin{figure}[h]
\begin{picture}(400,140)

\LongArrow(50,30)(350,30)
\Line(50,30)(50,40)
\Line(150,30)(150,40)
\Line(250,30)(250,120)
\Line(240,60)(250,60)
\Line(240,90)(250,90)
\Vertex(370,30)1

\Text(40,48)[l]{$\eta_2=0$}
\Text(115,48)[l]{$\eta_1=\eta$}
\Text(176,60)[l]{$z_1$}
\Text(150,90)[l]{$z_2$}
\Text(140,128)[l]{$\xi_1=1$}
\Text(236,48)[l]{$\xi_2=\infty$}

\end{picture}
\caption[]{Parametrization (I) of the world sheet of diagram (b)}  
\label{abbparambI}
\end{figure}
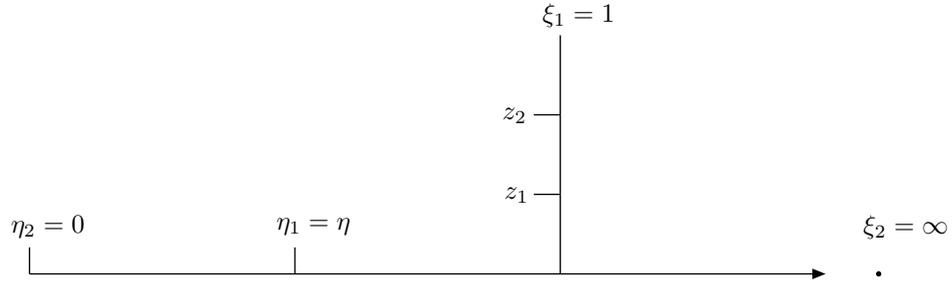 
    
\begin{figure}[h]
\begin{picture}(400,140)

\LongArrow(50,30)(350,30)
\Line(50,30)(50,40)
\Line(250,30)(250,40)
\Line(150,30)(150,120)
\Line(140,60)(150,60)
\Line(140,90)(150,90)
\Vertex(370,30)1

\Text(40,48)[l]{$\eta_2=0$}
\Text(115,128)[l]{$\eta_1=\eta$}
\Text(76,60)[l]{$z_1$}
\Text(50,90)[l]{$z_2$}
\Text(140,48)[l]{$\xi_1=1$}
\Text(236,48)[l]{$\xi_2=\infty$}

\end{picture}
\caption[]{Parametrization (II) of the world sheet of diagram (b)}  
\label{abbparambII}
\end{figure}
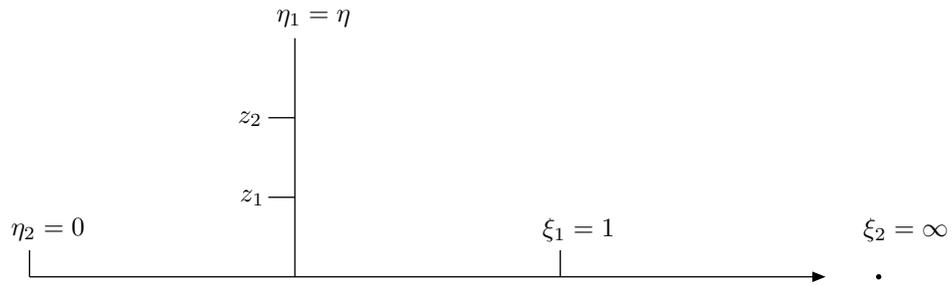 
\begin{figure}[h]
\begin{picture}(400,140)

\LongArrow(50,30)(350,30)
\Line(50,30)(50,40)
\Line(250,30)(250,90)
\Line(150,30)(150,90)
\Line(140,60)(150,60)
\Line(240,60)(250,60)
\Vertex(370,30)1

\Text(40,48)[l]{$\eta_2=0$}
\Text(115,98)[l]{$\eta_1=\eta$}
\Text(76,60)[l]{$z_1$}
\Text(150,60)[l]{$z_2$}
\Text(140,98)[l]{$\xi_1=1$}
\Text(236,48)[l]{$\xi_2=\infty$}

\end{picture}
\caption[]{Parametrization (III) of the world sheet of diagram (b)}  
\label{abbparamaI}
\end{figure}
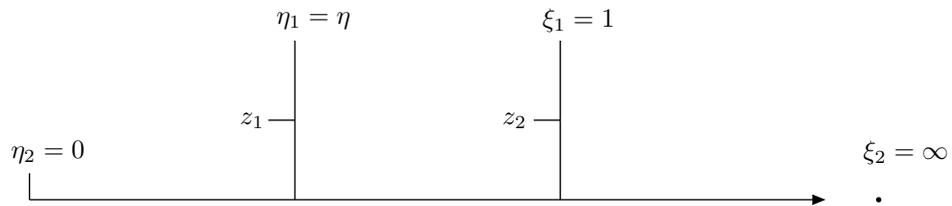     

\clearpage

\begin{table}[h]
\caption{Translation of the moduli into sewing parameters}
\label{tabparazwloanh}
\begin{center}
\begin{tabular}{llccc}
\vspace{0.2cm}
Diagram & Parametrization & $z_1$ & $z_2$ & $\eta$ \\
\vspace{0.2cm}
(a) & (I) & $A_1$ & $1-A_3$ & $A_1A_2$ \\
\vspace{0.2cm}
(a) & (II) & $A_1A_2$ & $1-A_3$ & $A_1$ \\
\vspace{0.2cm}
(a) & (III) & $A_1$ & $A_1A_2(1-A_3)$ & $A_1A_2$ \\
\vspace{0.2cm}
(a) & (IV) & $A_1A_2$ & $A_1(1-A_3)$ & $A_1$ \\
\vspace{0.2cm}
(b) & (I) & $1-A_2$ & $1-A_2A_3$ & $A_1$ \\
\vspace{0.2cm}
(b) & (II) & $A_1(1-A_2)$ & $A_1(1-A_2A_3)$ & $A_1$ \\
\vspace{0.2cm}
(b) & (III) & $A_1(1-A_2)$ & $1-A_3$ & $A_1$
\end{tabular}
\end{center}
\end{table}

\begin{table}[h]
\caption{Contributions to Feynman diagrams}
\label{tabbeitragzwloanh}
\begin{center}
\begin{tabular}{llc}
\vspace{0.2cm}
Diagram & Parametrization & Coefficient \\
\vspace{0.2cm}
(a) & (I) & $ \frac{25}{6} \eps^{-2} + \frac{181}{9} \eps^{-1} $ \\
\vspace{0.2cm}
(a) & (II) & $ \frac{121}{36} \eps^{-2} + \frac{3721}{216} \eps^{-1} $ \\
\vspace{0.2cm}
(a) & (III) & $ \frac{41}{6} \eps^{-2} + \frac{2327}{72} \eps^{-1} $ \\
\vspace{0.2cm}
(a) & (IV) & $ \frac{91}{36} \eps^{-2} + \frac{3199}{216} \eps^{-1} $ \\
\vspace{0.2cm}
(b) & (I) & $ \frac{293}{144} \eps^{-2} + \frac{26299}{4320} \eps^{-1} $ \\
\vspace{0.2cm}
(b) & (II) & $ \frac{521}{144} \eps^{-2} + \frac{54799}{4320}
\eps^{-1} $ \\
\vspace{0.2cm}
(b) & (III) & $ -\frac{13}{12} \eps^{-2} - \frac{369}{80} \eps^{-1} $
\end{tabular}
\end{center}
\end{table}

\clearpage

\section{Useful formulas and functions}
\label{formeln}

\setcounter{equation}{0}
\renewcommand{\theequation}{D.\arabic{equation}}

Formulas used to compute SPT integrals \cite{GRAD}:
\beqn \label{fsint1}
\int_0^\infty{\frac{x^{\mu-1}}{(\beta x+1)^\nu} dx} &=& \beta^{-\mu}
B(\mu,\nu-\mu), \\ 
& & \left[ \Re(\nu) > \Re(\mu) >0,\ \vert \arg(\beta)
\vert < \pi \right], \non 
\label{fsint2} 
\int_0^\infty{x^{\nu-1}(\beta+x)^{-\mu}(\gamma+x)^{-\rho} dx} &=& 
\beta^{-\mu} \gamma^{\nu-\rho} B(\nu, \mu-\nu+\rho) \\
& & \times {_2}F_1(\mu,\nu;\mu+\rho;1-\gamma/\beta), \non 
& & \left[ \Re(\nu) >0,\ \Re(\mu) > \Re(\nu-\rho),\right. \non
& & \left.  \vert \arg(\beta) \vert < \pi,\ \vert \arg(\gamma) \vert <
\pi \right], \non    
\label{fsint3}
\int_0^1{x^{\rho-1}(1-x)^{\sigma-1} {_2}F_1(\alpha,\beta;\gamma;x) dx} 
&=& B(\rho,\sigma) \  
{_3}F_2(\alpha,\beta,\rho;\gamma,\rho+\sigma;1), \\ 
& & \left[ \Re(\rho) >0,\ \Re(\sigma) >0,\
\Re(\gamma+\sigma-\alpha-\beta) >0 \right]. \nonumber 
\eeqn
The Gamma function:
\beqn \label{gammafunktion}
\Gamma (z) &=& \int_0^\infty{dt\ e^{-t} t^{z-1} }, \quad \left[ 
\Re (z) > 0 \right] .
\eeqn
The Euler Beta function:
\beqn \label{betafunktion}
B(\mu,\nu) &=& \frac{\Gamma(\mu) \Gamma(\nu)}{\Gamma(\mu+\nu)} \\
&=& \int_0^1{x^{\nu-1}(1-x)^{\mu-1}} dx, \quad \left[ \Re(\nu),
\Re(\mu)>0 \right]. \nonumber 
\eeqn
The generalized hypergeometric series:
\beqn \label{hypergeom}
{_p}F_q(\alpha_1,...,\alpha_p;\beta_1,...,\beta_q;z) = \sum_{k=0}^\infty{
\frac{(\alpha_1)_k \cdots (\alpha_p)_k}{(\beta_1)_k \cdots
  (\beta_q)_k} \frac{z^k}{k!} }, 
\eeqn
where $(\alpha)_k \equiv \alpha(\alpha+1)\cdots (\alpha+k-1)$ and
$(\alpha)_0 \equiv 1$. For ${_2}F_1$ there exists the integral 
representation \cite{GRAD}:
\beqn
{_2}F_1(\alpha,\beta;\gamma;z) &=& \frac{1}{B(\beta,\gamma-\beta)}
\int_{0}^1{t^{\beta-1}(1-t)^{\gamma-\beta-1}(1-tz)^{-\alpha} dt } ,\\
& & \left[ \Re(\gamma)> \Re(\beta) >0 \right] .\nonumber
\eeqn
For the integration of Gaussian integrals after applying the Schwinger trick 
to Feynman propagators 
\beqn \label{gaussint}
& & \int_{-\infty}^\infty{\frac{d^dpd^dk}{(2\pi)^{2d}} \exp \left( -\alpha
  p^2 -\beta k^2 -\delta q^2 +2\gamma pk +2xpq +2ykq \right) } = \\
& & \left( \alpha \beta -\gamma^2 \right) ^{-d/2} \exp \left( -q^2
  \frac{\alpha \delta \beta-\delta \gamma^2 -2xy \gamma -x^2 \beta
    -y^2 \alpha}{\alpha \beta-\gamma^2} \right) . \nonumber
\eeqn 
The Riemann Zeta function:
\beqn \label{zetafunktion}
\zeta(z) = \sum_{n=1}^\infty{n^{-z}}, \quad \left[ \Re (z)>1 \right].
\eeqn

\end{appendix}

\end{document}